\documentclass[12pt]{article}
\usepackage{amsmath}
\DeclareMathOperator{\logit}{logit}
\usepackage{amsthm}
\usepackage{graphicx}
\usepackage{enumitem} 
\usepackage{natbib}
\usepackage{url} 

\usepackage{tikz}
\usepackage{float}
\usepackage{amssymb}
\usepackage{hyperref}
\usepackage{bbm} 
\usepackage{dsfont}
\usepackage{booktabs}
\usepackage{subcaption}
\newcommand{\E}{\mathbb{E}}
\usepackage{caption}
\usepackage{longtable}
\usepackage[section]{placeins}
\usepackage{multibib}
\newcites{Supp}{References}

\DeclareMathOperator{\expit}{expit}
\newcommand{\blind}{0}
\newcommand{\independent}{\perp\!\!\!\perp} 
\newcommand{\notindependent}{ \not\!\perp\!\!\!\perp}
\newcommand{\bP}{\mathbb{P}}
\newcommand{\bE}{\mathbb{E}}

\newtheoremstyle{customMCAR}
  {}
  {}
  {\itshape}
  {}
  {\bfseries}
  {}
  {.5em}
  {\thmname{#1}\thmnumber{ #2}\textbf{ #3}}

\theoremstyle{customMCAR}

\newtheorem{assumption}{Assumption}
\newtheorem{theorem}{Theorem}

\newtheorem{result}{Result}
\newtheorem{proposition}{Proposition}
\newtheorem{remark}{Remark}

\begin{document}

\def\spacingset#1{\renewcommand{\baselinestretch}%
{#1}\small\normalsize} \spacingset{1}


\if0\blind
{
  \title{\bf 
  

  Identification and estimation of the conditional average treatment effect with nonignorable missing covariates, treatment, and outcome
  
  }
  
\author{%
  Shuozhi Zuo\thanks{Department of Statistics, University of Michigan, Ann Arbor}%
  \and
  Yixin Wang\thanks{Department of Statistics, University of Michigan, Ann Arbor}%
  \and
  Fan Yang\thanks{Yau Mathematical Sciences Center, Tsinghua University; Yanqi Lake Beijing Institute of Mathematical Sciences and Applications. Corresponding author: \texttt{yangfan1987@tsinghua.edu.cn}}%
}
    \date{}
  \maketitle
} \fi

\if1\blind
{
  \bigskip
  \bigskip
  \bigskip
  \medskip
} \fi
\bigskip
\begin{abstract}
Treatment effect heterogeneity is central to policy evaluation, social science, and precision medicine, where interventions can affect individuals differently. In observational studies, covariates, treatment, and outcomes are often only partially observed. When missingness depends on unobserved values (missing not at random; MNAR), standard methods can yield biased estimates of the conditional average treatment effect (CATE). This paper establishes nonparametric identification of the CATE under multivariate MNAR mechanisms that allow covariates, treatment, and outcomes to be MNAR. It also develops nonparametric and parametric estimators and proposes a sensitivity analysis framework for assessing robustness to violations of the missingness assumptions.
\end{abstract}

\noindent%
\vfill
\newpage
\spacingset{1.9} 
\section{Introduction}\label{sec:intro}

In many observational studies, researchers seek to understand how treatment effects vary across individuals. This heterogeneity is captured by the conditional average treatment effect (CATE), defined as the expected treatment effect given covariates. Accurate CATE estimation is essential for individualized decisions in both policy and medicine. In practice, however, covariates, treatment, and outcomes are often only partially observed. Missingness may depend on a variable's unobserved value as well as on other variables or missingness indicators, yielding complex multivariate missing not at random (MNAR) patterns.

Standard approaches for handling missing data, such as complete-case analysis (CCA), multiple imputation (MI) \citep{rubin1987multiple}, and inverse probability weighting (IPW) \citep{seaman2013review}, rely on strong assumptions about the missingness mechanism. CCA assumes missing completely at random (MCAR), meaning that missingness is independent of all variables. MI and IPW typically assume missing at random (MAR), under which missingness is independent of the unobservables conditional on the observables. In contrast, MNAR allows missingness to depend on the unobservables even conditional on all observables. Existing practical guidance \citep{lee2021framework} does not cover identification and estimation under multivariate MNAR mechanisms.

Across disciplines, reviews show that missing data are common but often underreported or handled by discarding incomplete cases. In the social sciences, \citet{berchtold2019treatment} reviewed quantitative articles published in 2017 across six leading journals and found that 69.5\% (105/151) of quantitative articles included missing data, yet only 44.4\% clearly reported its presence; most studies relied on discarding incomplete cases. Similar patterns also appear in education and psychology \citep{dong2013principled}. In epidemiology, \citet{mainzer2024gaps} reviewed 130 observational studies published between 2019 and 2021 in five leading journals. The review reported that 88\% had missing data in multiple variables, and only 34\% of studies stated any assumption about the missing-data mechanism.


The National Job Corps Study (NJCS) \citep{schochet2006national} illustrates the challenges that motivate this work. The goal is to characterize how the effect of obtaining a credential on earnings varies across participants. Several key variables exhibit substantial nonresponse, including covariates such as prior-year earnings and arrest history, the treatment of credential attainment, and the outcome of earnings, with missingness rates ranging between 7\% and 21\%. Because several measures are sensitive and collected by survey, selective nonresponse is likely. In particular, reporting of arrest history, prior earnings, credential status, and earnings may depend on the values themselves. Missingness may also depend on other partially observed variables. For example, baseline characteristics can affect follow-up participation, so later treatment and outcome missingness can depend on baseline values that are not always observed. Together, these features make MNAR a central concern in the NJCS.

\subsection{Related work}

Directed acyclic graphs (DAGs) encode dependencies among variables and missingness indicators, with directed edges representing potential causal relationships. DAGs make assumptions explicit and clarify which aspects of the data structure are restricted. 
Early work by \citet{fay1986causal}, \citet{ma2003identification}, \citet{mohan2014graphical}, and \citet{mohan2021graphical} used DAGs to derive identification results under a range of missingness patterns and clarified when parameters can be learned from incomplete data. Despite this progress, there is still no general algorithm that determines from a DAG whether a given parameter is identifiable from the observable data.

Previous work on MNAR in observational studies has largely focused on identifying the average treatment effect (ATE) under either restrictive missingness mechanisms or by limiting missingness to only a few variables. When only covariates are missing, \citet{blake2020estimating} use outcome regression augmented with missingness indicators without restricting the missingness mechanism. However, their approach requires a modified strong ignorability assumption \citep{rosenbaum1984reducing} that allows a covariate to confound treatment and outcome only when observed, an assumption difficult to justify in practice. To avoid this assumption, other work imposes structure on the missingness mechanism. Under outcome-independent covariate missingness, where missingness is independent of the outcome conditional on covariates and treatment,  \citet{yang2019causal}, \citet{guan2024unified}, and \citet{sun2021semiparametric} establish ATE identification under a completeness condition. \citet{sun2025identification} consider treatment-independent covariate missingness and derive identification results for the ATE under parametric models when missingness exists in a single covariate. When both treatment and covariates have missing data, \citet{wen2025incomplete} study ATE identification requiring treatment missingness to be independent of the outcome conditional on treatment and covariates, while covariate missingness must be independent of both the outcome and its own unobserved value given treatment and observed variables.

Identification under MNAR becomes more challenging when the outcome is also missing. From a distribution recovery perspective,  
\citet{moreno2018canonical} show that identification often fails when variables are allowed to affect their own missingness, i.e., self-censoring. From a regression perspective, \citet{hughes2019accounting} demonstrate that the treatment coefficient can be recovered by CCA when the missingness of all variables is conditionally independent of the outcome.
\citet{landsiedel2025causal} study ATE identification also under the constraint that outcome missingness does not depend on the outcome itself. \citet{chen2023selfcensoring} directly address outcome self-censoring in ATE identification using treatment as a shadow variable but do not allow covariates or treatment to be missing.

In observational studies, most of the literature on missing data targets the ATE, which typically requires recovering the joint distribution of treatment, covariates and outcome, and thus also identifies the CATE when achievable. Our focus is complementary: when several variables are MNAR, recovering the full joint distribution may require strong missingness assumptions, while the CATE can remain identifiable under weaker conditions because it depends only on the conditional outcome distribution. We illustrate this distinction in Section~\ref{app:counterexample-general1} of the supplementary material through a counterexample where the CATE is identifiable but the ATE is not. Methodological work that directly targets the CATE under missingness remains limited. \citet{kuzmanovic2023estimating} study the estimation of CATE when treatment is MAR. \citet{ding2014identifiability} establish identification of the CATE with MNAR covariates in randomized studies with fully observed treatment and outcome. A key distinction from randomized settings is that in observational studies, covariates also act as confounders, further complicating identification.

\subsection{Our contributions}


This paper studies identification and estimation of the CATE under general multivariate MNAR mechanisms, allowing covariates, treatment, and outcome to be partially observed, including self-censoring. In prospective studies where covariates and treatment are measured before the outcome, the framework allows general MNAR mechanisms for covariates and treatment while imposing restrictions only on the outcome-missingness process. We study and establish nonparametric identification of the CATE under three outcome-missingness mechanisms: outcome-independent, treatment-independent, and covariate-independent mechanisms. Counterexamples in Section~\ref{app:counterexample-general} of the supplementary material show that these restrictions do not generally identify the ATE and that the CATE becomes unidentifiable under more general outcome-missingness mechanisms. The paper also develops corresponding nonparametric and parametric estimators. Because each baseline assumption removes exactly one pathway into outcome missingness, sensitivity analysis can be implemented by reintroducing that pathway through a single-parameter extension.

\subsection{Organization of the paper}

Section~\ref{sec:notation} introduces the notations, definitions, and causal assumptions.  
Section~\ref{sec:missingness} presents the missingness mechanisms and corresponding identification results.  
Section~\ref{sec:estimation} describes the proposed estimation approaches.  
Section~\ref{sec:sim} reports simulation studies comparing estimators under different missingness assumptions.  
Section~\ref{sec:sensitivity} presents the NJCS application and the sensitivity analysis.  
Section~\ref{sec:discussion} concludes.

\section{Notation, definition, and causal assumptions}\label{sec:notation}

Consider a sample of size $n$ consisting of independent and identically distributed observations drawn from a superpopulation.  
Let $X$, $T$, and $Y$ denote the covariates, treatment, and outcome, with supports $\mathcal{X}$, $\mathcal{T}$, and $\mathcal{Y}$, respectively.
Let $R^Y$ be the response indicator such that $R^Y=1$ if $Y$ is observed and $R^Y=0$ otherwise.  
Similarly, let $R^T$ and $R^X$ denote the response indicators for $T$ and $X$, respectively.  
In general, $X$ is a vector of covariates, in which case we write
$X=(X_1,\ldots,X_p)$ and $R^X=(R^{X_1},\ldots,R^{X_p})$ for the corresponding componentwise response indicators. For notational simplicity, we use the
scalar notation $X$ and $R^X$ throughout, and all identification results and estimators extend directly to the multivariate case. We adopt the potential outcomes framework, where each unit has potential outcomes
$\{Y(t): t \in \mathcal{T}\}$ for all treatment levels.
The parameter of interest is the CATE comparing a pair of treatment levels $t_1\in \mathcal{T}$ and $t_0\in \mathcal{T}$,
\[
\tau_{t_1,t_0}(x)=\E\{Y(t_1)-Y(t_0)\mid X=x\},
\]
defined for each covariate value $x\in \mathcal{X}$.

We impose the following causal assumptions: (i) the stable unit treatment value assumption (SUTVA), which requires consistency and no interference, so that the observed outcome satisfies $Y=Y(T)$;
(ii) strong ignorability, $\{Y(t): t\in\mathcal{T}\}\independent T\mid X$; and
(iii) positivity, $\bP(T=t\mid X=x)>0$ for all $t\in \mathcal{T}$ and all $x\in\mathcal{X}$.
Under these assumptions,
\[
\E\{Y(t)\mid X=x\}=\E\{Y\mid T=t,X=x\},
\]
so identification of $\bP(Y\mid T,X)$ suffices to identify $\tau_{t_1,t_0}(x)$.

Two of our identification results rely on a completeness condition. A function $f(A,B)$ is said to be \emph{complete in $B$} if, for any square-integrable function $g$,
\[
\int g(A)\,f(A,B)\,d\nu(A)=0
\quad \Longrightarrow \quad
g(A)=0 \ \text{a.s.},
\]
where $\nu(\cdot)$ denotes the Lebesgue measure for continuous $A$ and the counting measure for discrete $A$.

\section{Missingness mechanisms and identification}\label{sec:missingness}

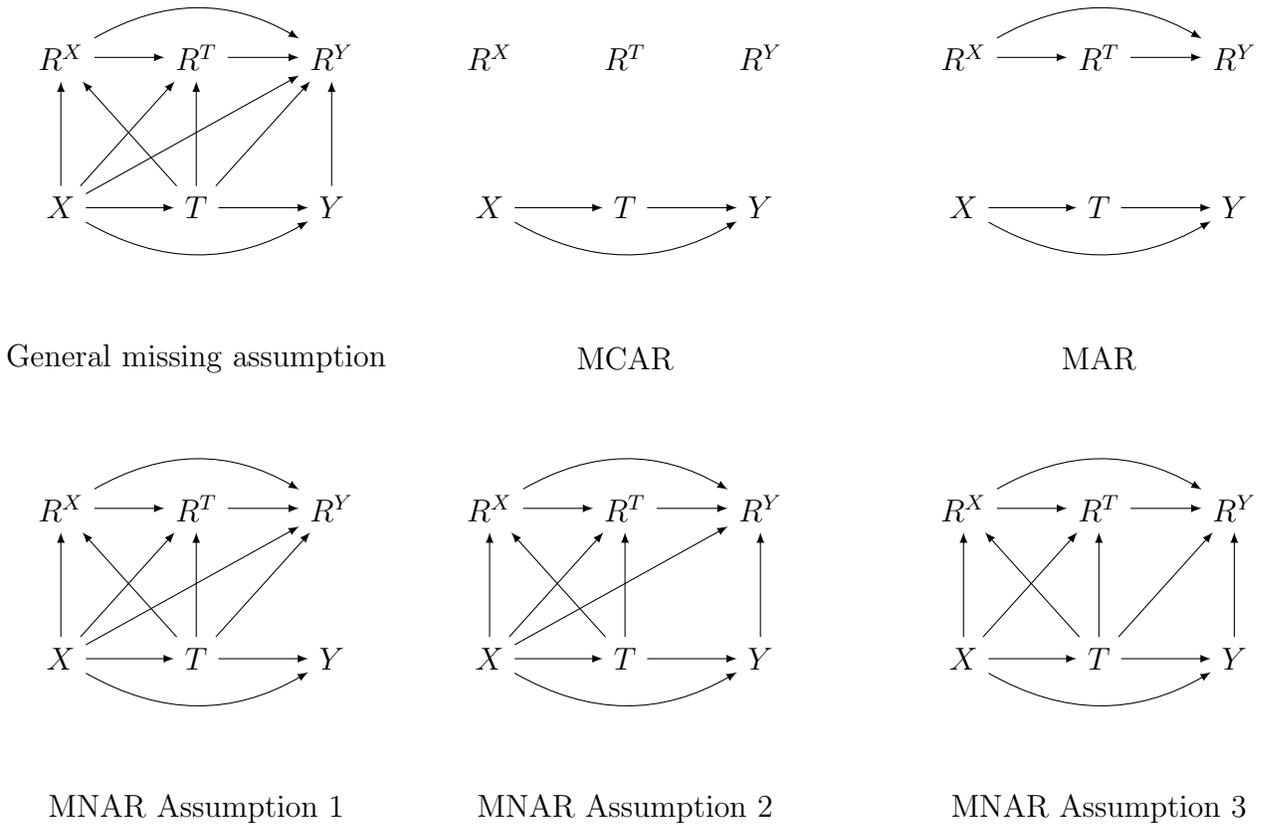
\begin{figure}[ht]
\centering
\begin{tikzpicture}

    \node (x)  at (-9,6) {$X$};
    \node (t)  at (-7.2,6) {$T$};
    \node (y)  at (-5.4,6) {$Y$};
    \node (rx) at (-9,8) {$R^X$};
    \node (rt) at (-7.2,8) {$R^T$};
    \node (ry) at (-5.4,8) {$R^Y$};
    \node (a)  at (-7.2,4) {General missing assumption};
    \path[-latex] (t) edge (y);
    \path[-latex] (x) edge (t);
    \path[-latex] (x) edge [bend right] (y);
    \path[-latex] (x) edge (rx);
    \path[-latex] (x) edge (rt);
    \path[-latex] (x) edge (ry);
    \path[-latex] (t) edge (rx);
    \path[-latex] (t) edge (rt);
    \path[-latex] (t) edge (ry);
    \path[-latex] (y) edge (ry);
    \path[-latex] (rx) edge (rt);
    \path[-latex] (rx) edge [bend left] (ry);
    \path[-latex] (rt) edge (ry);

    \node (x)  at (-3.3,6) {$X$};
    \node (t)  at (-1.5,6) {$T$};
    \node (y)  at (0.3,6) {$Y$};
    \node (rx) at (-3.3,8) {$R^X$};
    \node (rt) at (-1.5,8) {$R^T$};
    \node (ry) at (0.3,8) {$R^Y$};
    \node (a)  at (-1.5,4) {MCAR};
        \path[-latex] (t) edge (y);
    \path[-latex] (x) edge (t);
    \path[-latex] (x) edge [bend right] (y);

    \node (x)  at (3.0,6) {$X$};
    \node (t)  at (4.8,6) {$T$};
    \node (y)  at (6.6,6) {$Y$};
    \node (rx) at (3.0,8) {$R^X$};
    \node (rt) at (4.8,8) {$R^T$};
    \node (ry) at (6.6,8) {$R^Y$};
    \node (a)  at (4.8,4) {MAR};
        \path[-latex] (t) edge (y);
    \path[-latex] (x) edge (t);
    \path[-latex] (x) edge [bend right] (y);
    \path[-latex] (rx) edge (rt);
    \path[-latex] (rx) edge [bend left] (ry);
    \path[-latex] (rt) edge (ry);
    
    \node (x)  at (-9,0) {$X$};
    \node (t)  at (-7.2,0) {$T$};
    \node (y)  at (-5.4,0) {$Y$};
    \node (rx) at (-9,2) {$R^X$};
    \node (rt) at (-7.2,2) {$R^T$};
    \node (ry) at (-5.4,2) {$R^Y$};
    \node (a)  at (-7.2,-2) {MNAR Assumption \ref{ass1}};
    \path[-latex] (t) edge (y);
    \path[-latex] (x) edge (t);
    \path[-latex] (x) edge [bend right] (y);
    \path[-latex] (x) edge (ry);
    \path[-latex] (t) edge (ry);
    \path[-latex] (rx) edge [bend left] (ry);
    \path[-latex] (rt) edge (ry);
    \path[-latex] (x) edge (rx);
    \path[-latex] (x) edge (rt);
    \path[-latex] (t) edge (rx);
    \path[-latex] (t) edge (rt);
    \path[-latex] (rx) edge (rt);

    \node (x)  at (-3.3,0) {$X$};
    \node (t)  at (-1.5,0) {$T$};
    \node (y)  at (0.3,0) {$Y$};
    \node (rx) at (-3.3,2) {$R^X$};
    \node (rt) at (-1.5,2) {$R^T$};
    \node (ry) at (0.3,2) {$R^Y$};
    \node (a)  at (-1.5,-2) {MNAR Assumption \ref{ass2}};
    \path[-latex] (t) edge (y);
    \path[-latex] (x) edge (t);
    \path[-latex] (x) edge [bend right] (y);
    \path[-latex] (x) edge (ry);
    \path[-latex] (y) edge (ry);
    \path[-latex] (rx) edge [bend left] (ry);
    \path[-latex] (rt) edge (ry);
    \path[-latex] (x) edge (rx);
    \path[-latex] (x) edge (rt);
    \path[-latex] (t) edge (rx);
    \path[-latex] (t) edge (rt);
    \path[-latex] (rx) edge (rt);

    \node (x)  at (3.0,0) {$X$};
    \node (t)  at (4.8,0) {$T$};
    \node (y)  at (6.6,0) {$Y$};
    \node (rx) at (3.0,2) {$R^X$};
    \node (rt) at (4.8,2) {$R^T$};
    \node (ry) at (6.6,2) {$R^Y$};
    \node (a)  at (4.8,-2) {MNAR Assumption \ref{ass3}};
    \path[-latex] (t) edge (y);
    \path[-latex] (x) edge (t);
    \path[-latex] (x) edge [bend right] (y);
    \path[-latex] (t) edge (ry);
    \path[-latex] (y) edge (ry);
    \path[-latex] (rx) edge [bend left] (ry);
    \path[-latex] (rt) edge (ry);
    \path[-latex] (x) edge (rx);
    \path[-latex] (x) edge (rt);
    \path[-latex] (t) edge (rx);
    \path[-latex] (t) edge (rt);
    \path[-latex] (rx) edge (rt);

\end{tikzpicture}
\caption{Each MNAR assumption restricts exactly one arrow into $R^Y$ relative to the general missingness DAG. Top row: general missingness, MCAR, and MAR. Bottom row: MNAR Assumptions~\ref{ass1}--\ref{ass3}.}
\label{fig:missingmodel}
\end{figure}

We focus on prospective studies in which $X$ and $T$ are measured before $Y$, reflecting the natural temporal ordering commonly observed in applied research. Accordingly, it is reasonable to assume that future outcomes do not influence the past events; that is, $(Y,R^Y)$ are assumed not to affect $(X,T,R^X,R^T)$. Typically, covariates $X$ are measured before treatment $T$ is administered; however, in settings where $X$ and $T$ are obtained within the same survey, both variables may influence $R^X$ and $R^T$, and our framework accommodates this possibility. Under this setup, the DAGs in Figure~\ref{fig:missingmodel} illustrate the most general missing data assumption, as well as the MCAR, MAR, and MNAR Assumptions~\ref{ass1}--\ref{ass3}. DAGs encode missingness assumptions as conditional independence restrictions at the variable level. The MAR assumption represented by DAGs differs from the form typically used in MI, where missingness may depend on a variable only when it is observed. As noted by \citet{mealli2015clarifying}, this MI formulation is not a conditional independence assumption and becomes difficult to interpret or justify when multiple variables are subject to missingness. For this reason, we adopt the MAR assumption formalized via DAGs \citep{mohan2014graphical}.

In the first row of Figure~\ref{fig:missingmodel}, the most general missingness assumption allows the missingness
indicators $(R^X,R^T,R^Y)$ to depend arbitrarily on $(X,T,Y)$ and on one another, subject only to the temporal
restriction described above. Under this general mechanism, the observed data do not, in general, identify $\tau_{t_1,t_0}(x)$; a simple counterexample is provided in Section~\ref{app:counterexample-general2} of the supplementary material. Under MCAR, each missingness
indicator is independent of all other variables. Under MAR, the missingness indicators may depend on one
another but cannot depend on the partially observed variables $(X,T,Y)$. The three MNAR mechanisms in the second row are derived from the most general assumption by removing a single arrow into $R^Y$, thereby imposing additional restrictions on the outcome-missingness process as formalized in Assumptions~\ref{ass1}--\ref{ass3}.

\begin{assumption}
$R^X\independent Y\mid (X,T)$, $R^T\independent Y\mid (X,T,R^X)$, $R^Y\independent Y\mid (X,T,R^X,R^T)$.\label{ass1}
\end{assumption}

\begin{assumption}
$R^X\independent Y\mid (X,T)$, $R^T\independent Y\mid (X,T,R^X)$, $R^Y\independent T\mid (X,Y,R^X,R^T)$.\label{ass2}
\end{assumption}

\begin{assumption}
$R^X\independent Y\mid (X,T)$, $R^T\independent Y\mid (X,T,R^X)$, $R^Y\independent X\mid (T,Y,R^X,R^T)$.\label{ass3}
\end{assumption}

Assumptions~\ref{ass1}--\ref{ass3} maintain the general missingness structure for $X$ and $T$. Specifically, we allow $R^X$ to depend on $(X, T)$ and $R^T$ to depend on $(X, T, R^X)$. The differences among Assumptions~\ref{ass1}--\ref{ass3} lie in the specification of the missingness mechanism for $Y$. Under Assumption~1, $R^Y$ may depend on $(X, T, R^X, R^T)$ but is conditionally independent of $Y$. Under Assumption~2, $R^Y$ may depend on $(X, Y, R^X, R^T)$ but is conditionally independent of $T$. Under Assumption~3, $R^Y$ may depend on $(T, Y, R^X, R^T)$ but is conditionally independent of $X$. While identification of the CATE under Assumption~\ref{ass1} has been discussed previously \citep{moreno2018canonical}, Assumptions~\ref{ass2} and~\ref{ass3} enable new identification results for the CATE in settings where the outcome may be self-censored.

The following theorems present the conditions required for nonparametric identification of $\tau_{t_1,t_0}(x)$ under MNAR Assumptions~\ref{ass1}--\ref{ass3}.

\begin{theorem}\label{the1}
Under the causal assumptions (i) to (iii) and Assumption \ref{ass1}, if $\bP(R^Y=1,R^T=1,R^X=1\mid X=x,T=t)>0$ for all $x,t$, then $\bP(Y\mid T,X)$ is identifiable, and therefore, $\tau_{t_1,t_0}(x)$ is identifiable.
\end{theorem}

Assumption~\ref{ass1} rules out outcome self-censoring, implying that the outcome does not influence its own missingness. This corresponds to case (m-DAG~E) in the framework of \citet{moreno2018canonical}, under which the conditional outcome distribution is identifiable from the complete cases because $\bP(Y|X,T) = \bP(Y|X, T, R^X=1, R^T=1, R^Y=1)$.
MCAR and MAR are both special cases of Assumption~\ref{ass1}. Consequently, when the missingness in $Y$ is conditionally independent of $Y$ itself, CCA yields consistent estimates of $\bP(Y\mid T,X)$ and hence of $\tau_{t_1,t_0}(x)$.

\begin{theorem}\label{the2}
Under the causal assumptions (i) to (iii) and Assumption~\ref{ass2}, suppose
$\bP(R^Y=1\mid X=x,Y=y,R^X=1,R^T=1)>0$ for all $x,y$ and
$\bP(R^T=1,R^X=1\mid T=t,X=x)>0$ for all $t,x$.
\begin{enumerate}[label=(\roman*)]
\item
If $\bP(Y,R^Y=1 \mid T,X=x,R^T=1,R^X=1)$ is complete in $T$ for $x\in\mathcal{X}$,
then $\bP(Y\mid T,X=x)$ is identifiable, and hence $\tau_{t_1,t_0}(x)$ is identifiable.

\item
For any $x$ such that $Y \independent T \mid X=x$, the null stratum-specific effect
$\tau_{t_1,t_0}(x)=0$ is identifiable, although the completeness condition in part (i) fails at that $x$.
\end{enumerate}
\end{theorem}

Part (i) of Theorem~\ref{the2} relies on a completeness condition, an assumption widely used in nonparametric identification problems. In MNAR settings where a partially observed variable may influence its own missingness  \citep{miao2024shadow, yang2019causal, li2022self, zuo2025mediation}, i.e., self-censoring, the completeness condition leverages a shadow variable \citep{miao2024shadow} that is associated with the partially observed variable but, conditional on other observed information, does not affect its missingness. Completeness then guarantees that the missingness mechanism is uniquely identified from the observed data. Sufficient conditions for completeness are provided in \citet{dhaultfoeuille2010new}. This property also holds for many commonly used parametric families under some assumptions, including exponential-family models \citep{newey2003instrumental} and certain location-scale families \citep{hu2018nonparametric}. We provide simple examples tailored to Theorems~\ref{the2} and \ref{the3} in Section~\ref{app:param-ex} of the supplementary material.

To build intuition for part (i) and to illustrate how completeness facilitates identification, we consider a simple discrete example. Suppose that $T$ takes $J$ distinct values $\{t_1,\ldots,t_J\}$ and $Y$ takes $K$ distinct values $\{y_1,\ldots,y_K\}$. Because
\begin{align}
\bP(Y=y\mid X=x,T=t)
= \frac{\bP(Y=y,R^Y=1\mid X=x,T=t,R^X=1,R^T=1)}
         {\bP(R^Y=1\mid X=x,Y=y,R^X=1,R^T=1)},\nonumber
\end{align}
the conditional distribution $\bP(Y=y\mid X=x,T=t)$ is identifiable if $\bP(R^Y=1\mid X=x,Y=y,R^X=1,R^T=1)$ is identifiable. For each $x$, define
\[
\zeta_x(y_k)
= \frac{\bP(R^Y = 0 \mid X = x, Y = y_k, R^X = 1, R^T = 1)}
        {\bP(R^Y = 1 \mid X = x, Y = y_k, R^X = 1, R^T = 1)},
\qquad k = 1,\ldots,K,
\]
which represents the unknown odds that $Y$ is missing at each level $y_k$. For each treatment level $t_j$, we have the following identity
\begin{eqnarray*}
&&\bP(R^Y = 0 \mid T = t_j, X = x, R^T = 1, R^X = 1) \\&=& \sum_{k=1}^K
   \bP(Y = y_k, R^Y = 1 \mid T = t_j, X = x, R^T = 1, R^X = 1)\,
   \zeta_x(y_k).
\end{eqnarray*}
Thus, for each $x$, we obtain a system of $J$ linear equations in the $K$ unknowns
$\{\zeta_x(y_k)\}_{k=1}^K$. Let $\Theta_x$ denote the $J \times K$ matrix with entries
\[
\Theta_x(j,k) = \bP(Y = y_k, R^Y = 1 \mid T = t_j, X = x, R^T = 1, R^X = 1).
\]
To ensure a unique solution for $\zeta_x(y_k)$'s and hence for $\bP(R^Y=1\mid X=x,Y=y,R^X=1,R^T=1)=\{1+\zeta_x(y)\}^{-1}$, it suffices to impose the full-rank condition $\mathrm{rank}(\Theta_x)=K$. This is the completeness condition in the discrete case. This condition essentially requires that $J\geq K$ and $T \notindependent  Y\mid X$.

Part (ii) highlights an important case in which $Y\independent T\mid X=x$ holds for some $x$. In this situation, $T$ no longer serves as a shadow variable, and consequently the completeness condition fails at $X=x$. Therefore, the conditional distribution $\bP(Y\mid T,X=x)$ is no longer identifiable. Nevertheless, under Assumption~\ref{ass2}, we show that $\tau_{t_1,t_0}(x)$ remains identifiable and equals $0$. Intuitively, because $R^Y \independent T \mid (X,Y,R^X,R^T)$, the outcome-missingness mechanism is identical across treatment levels after conditioning on $(X,Y,R^X,R^T)$. Therefore, any differences in observed outcome distributions across treatment levels reflect differences in the underlying conditional outcome distribution $\bP(Y\mid T,X=x)$. When no such differences exist at $X=x$, $\tau_{t_1,t_0}(x)=0$ is correctly identified even though $\bP(Y\mid T,X=x)$ may not be recoverable.

Full proofs are provided in Section~S2 of the supplementary material.

\begin{theorem}\label{the3}
Under the causal assumptions (i) to (iii) and Assumption \ref{ass3}, if
$\bP(R^Y=1\mid T=t,Y=y,R^X=1,R^T=1)>0$ for all $t,y$,
$\bP(R^T=1,R^X=1\mid T=t,X=x)>0$ for all $t,x$, and
$\bP(Y,R^Y=1\mid T=t,X,R^T=1,R^X=1)$ is complete in $X$ for all $t$,
then $\bP(Y\mid T,X)$ is identifiable, and therefore, $\tau_{t_1,t_0}(x)$ is identifiable. 
\end{theorem}

\begin{remark}\label{rem:idX}
When $X$ is multivariate, we can relax Theorem \ref{the3} as follows. Write $X=(X^{\mathrm{id}},X^{\mathrm{c}})$, where $X^{\mathrm{id}}$ denotes the covariate component that satisfies (i) the modified conditional independence assumption in Assumption~\ref{ass3},
$R^Y \independent X^{\mathrm{id}} \mid (T,Y,R^{X^{\mathrm{id}}},R^{X^{\mathrm{c}}},R^T,X^{\mathrm{c}}),$
and (ii) the modified completeness condition required in Theorem~\ref{the3},
$\bP(Y,R^Y=1\mid T=t,X,R^T=1,R^X=1)$ is complete in $X^{\mathrm{id}}$ for all $t$.
Under these two modified conditions, the conclusions of Theorem \ref{the3} continue to hold. The estimation and inference strategies described in Section \ref{sec:estimation} can be straightforwardly extended under the modified assumptions.
\end{remark}

The intuition behind Theorem~\ref{the3} parallels part (i) of Theorem~\ref{the2}, with the roles of $T$ and $X$ reversed. Under Assumption~\ref{ass3}, $R^Y \independent X \mid (T,Y,R^X,R^T)$, implying that,  within each treatment stratum $T=t$, the outcome-missingness mechanism is invariant across values of $X$. The completeness condition ensures that $X$ provides sufficient variation relative to $Y$ (within each $T=t$) to identify this missingness mechanism, thereby allowing recovery of $\bP(Y\mid T,X)$ from the observed data.

\paragraph{Summary.} Theorems~\ref{the1}--\ref{the3} identify the CATE in prospective studies where $(X,T)$ are measured before $Y$, precluding $(Y,R^Y)$ from influencing the missingness of $X$ or $T$. Under this time ordering, we allow the most general MNAR mechanisms for $(X,T)$ and restrict only the outcome-missingness process, leading to three identifiable cases in Figure~\ref{fig:missingmodel}: (i) Assumption~\ref{ass1}, which rules out outcome self-censoring and identifies $\bP(Y\mid X,T)$ from complete cases; (ii) Assumption~\ref{ass2}, which allows outcome self-censoring but imposes $R^Y \perp\!\!\!\perp T \mid (X,Y,R^X,R^T)$; and (iii) Assumption~\ref{ass3}, which allows outcome self-censoring but imposes $R^Y \perp\!\!\!\perp X \mid (T,Y,R^X,R^T)$. Under Assumptions~\ref{ass2}--\ref{ass3}, identification additionally requires the corresponding completeness condition, except when 
$Y\independent T\mid X=x$ under Assumption~\ref{ass2}, in which case the CATE is identifiable and equals zero even without completeness.

These identification results imply that estimation and inference for the CATE should be conducted using analyses restricted to units with observed $(X,T)$, as illustrated in the sketch of the proof for Theorem \ref{the2}. An important distinction between identifying the CATE and the ATE is that the CATE depends only on $\bP(Y\mid X,T)$, and thus can remain identifiable even when the joint distribution of $(X, T)$ is not. As a result, the CATE can be identified under weaker and more realistic missingness assumptions than those required for the ATE. Assumptions~\ref{ass1}--\ref{ass3} do not generally identify the full-data joint law and therefore do not, in general, identify the ATE; we provide counterexamples in Section \ref{app:counterexample-general1} of the supplementary material. Importantly, even with correctly specified models for the joint distribution of $(X, T, Y, R^X, R^T, R^Y)$, the MLE based on the full-data likelihood may yield biased estimates of the CATE under the MNAR mechanisms in Assumptions~\ref{ass1}--\ref{ass3}, precisely because the joint distribution involves non-identifiable components $(X, T)$. This point is reflected in our simulation results.

The missingness mechanisms in Assumptions~\ref{ass1}--\ref{ass3} are highly flexible, as each excludes only a single pathway into $R^Y$ relative to the most general missing DAG. This structure makes sensitivity analysis particularly straightforward. In Section~\ref{sec:sensitivity}, we embed each baseline missingness model in a single-parameter family that reintroduces the excluded dependence (e.g., allowing $Y\to R^Y$, $T\to R^Y$, or $X\to R^Y$) and assess the robustness of the estimated CATE over a range of quantified departures from the baseline assumption.

\section{Estimation}\label{sec:estimation}

Under Assumption~\ref{ass1}, $\bP(Y|X, T) = \bP(Y|X, T, R^X=1, R^T=1, R^Y=1)$, the complete-case estimator is consistent for
$\bP(Y \mid X,T)$ and hence for $\tau_{t_1,t_0}(x)$. For example, a nonparametric estimator
can be obtained by flexibly regressing $Y$ on $(X,T)$ among complete cases
(e.g., via kernel-based or spline smoothing), yielding
$\widehat\mu(t,x)\approx \bE(Y\mid T=t,X=x)$ and
$\widehat\tau_{t_1,t_0}(x)=\widehat\mu(t_1,x)-\widehat\mu(t_0,x)$.
Alternatively, under a parametric outcome model $\bP_\beta(y\mid x,t)$, one can estimate
$\beta$ by maximizing the complete-case likelihood and then obtain
$\tau_{t_1,t_0}(x)$ via plug-in.

The remainder of this section focuses on estimation under Assumptions~\ref{ass2} and~\ref{ass3}. Under these two assumptions, $R^Y$ may depend on $Y$ itself, rendering the complete-case estimator generally biased due to selection. To address this, we develop both nonparametric and parametric estimators that implement the corresponding identification results: (i) a nonparametric series two-stage least squares (2SLS) estimator that solves the associated
Fredholm equation for the response-odds function via a sieve approximation and quadratic
regularization \citep{kress1999linear,newey2003instrumental,yang2019causal}; and
(ii) a parametric likelihood-based estimator computed via the expectation–maximization (EM) algorithm \citep{dempster1977maximum}.

\subsection{Nonparametric series 2SLS estimator under Theorem \ref{the2} or \ref{the3}}\label{subsec:np-tsls}

We implement the identification results in Theorems~\ref{the2} and~\ref{the3} by
(i) recovering the outcome response mechanism through an integral equation solved via a series 2SLS approach, and
(ii) using the estimated response odds to construct a selection-corrected estimator of
$\bE(Y\mid X,T)$ and consequently of $\tau_{t_1,t_0}(x)$. Based on the identification results, we restrict analysis to units with observed $(X,T)$, i.e., $\{i: R^X=R^T=1\}$. We present the construction under Theorem~\ref{the2};
an analogous procedure applies to Theorem~\ref{the3} after swapping the roles of $T$ and $X$.

Define the response odds function
$$
\zeta(x,y)
=
\frac{\bP(R^Y=0\mid X=x,Y=y,R^X=1,R^T=1)}{\bP(R^Y=1\mid X=x,Y=y,R^X=1,R^T=1)},
$$
and let $
\pi(x,y)
=\bP(R^Y=1\mid X=x,Y=y,R^X=1,R^T=1)
=\{1+\zeta(x,y)\}^{-1}.
$

The probabilities below satisfy the Fredholm equation
\begin{eqnarray}\label{eq:fredholm_ass2_generic}
&&\bP(R^Y=0\mid T=t,X=x,R^X=1,R^T=1)\nonumber\\&=&
\int \bP(Y=y,R^Y=1\mid T=t,X=x,R^X=1,R^T=1)\,\zeta(x,y)\,dy.
\end{eqnarray}

To operationalize \eqref{eq:fredholm_ass2_generic}, we approximate function $\zeta$ in a finite-dimensional
sieve space and enforce the integral equation through its sample analogue, yielding a
regularized least-squares problem. Let
$h(y)=(h_1(y),\ldots,h_J(y))^\top$ be a chosen sieve basis in $y$ (e.g., a Hermite-type envelope basis).
When $x$ is discrete, for each $x$ we use the approximation $\zeta(x,y)\approx h(y)^\top\beta(x)$, which leads to a linear
system $\widehat b_x \approx \widehat M_x\,\beta(x)$ obtained from the sample analogue of (\ref{eq:fredholm_ass2_generic}) over $t\in \mathcal{T}$.
We estimate $\beta(x)$ by minimizing the integral-equation residual subject to quadratic
regularization,
\[
\widehat\beta(x)
=
\arg\min_{\beta}\ \|\widehat b_x-\widehat M_x\beta\|^2
\quad\text{subject to}\quad
\beta^\top \Lambda \beta \le B,
\]
where $\Lambda$ is a positive definite penalty matrix that regularizes the sieve coefficients and
$B>0$ controls the regularization strength. We then set
$\widehat\zeta(x,y)=h(y)^\top\widehat\beta(x)$. When $X$ is continuous, we instead use a tensor-product sieve
$\zeta(x,y)\approx \phi(x,y)^\top\theta$ with $\phi(x,y)=g(x)\otimes h(y)$, where
$g(x)=(g_1(x),\ldots,g_{J_x}(x))^\top$ is a chosen sieve basis in $x$ (e.g., the same
Hermite-type envelope basis applied to a standardized $x$). Finally, we form weights $\widehat w_i = 1+\widehat\zeta(X_i,Y_i)$ and estimate $\bE(Y\mid X,T)$ via
weighted nonparametric regression on complete cases, i.e., units $\{i: R^X=R^T=R^Y=1\}$, yielding $\widehat\tau_{t_1,t_0}(x)$. Implementation details are provided in
Section~\ref{app:np2sls-impl} of the supplementary material.

As highlighted in part (ii) of Theorem~\ref{the2}, when $Y\independent T\mid X=x$, the completeness condition in part (i) of Theorem~\ref{the2} fails, rendering $\bP(Y\mid T,X=x)$ (and thus $\bE(Y\mid X=x,T)$) not identifiable. In this case, an estimator targeting $\bE(Y\mid X=x,T)$ may be biased, yet the bias in the estimated $\tau_{t_1,t_0}(x)$ is expected to be negligible, since $\tau_{t_1,t_0}(x)$ can still be correctly identified as $0$. In contrast, when the completeness condition holds (part (i) of Theorem \ref{the2}), the biases in both $\bE(Y\mid X=x,T)$ and $\tau_{t_1,t_0}(x)$ are expected to be small. These expectations are consistent with our simulation results.

For inference, we follow the bootstrap procedure recommended in \citet{yang2019causal} with all
tuning parameters fixed across bootstrap resamples.

\subsection{Parametric estimation under Theorem \ref{the2} or \ref{the3}}\label{subsec:EM}

The series 2SLS estimator directly implements the nonparametric identification results and is flexible, but it can be unstable due to the ill-posed nature of the integral equation and its sensitivity to tuning parameters. As a more stable alternative, we consider a parametric estimator implemented via the EM algorithm. This approach imposes structure on the outcome and missingness models and can yield smoother and more stable estimates when the parametric assumptions are adequate.

We again restrict our analysis to the subset of units with observed $(X,T)$.
We specify a parametric model for the outcome,
$
\bP_\beta(y\mid x,t),
$
for example a generalized linear model
\[
g\{\E_\beta(Y\mid x,t)\}=U(x,t)^\top\beta.
\]
Under Assumptions \ref{ass2} and \ref{ass3}, $\bP_\beta(y\mid x,t) =\bP_\beta(y\mid x,t, R^X=1, R^T=1)$.
We also specify a parametric model for the outcome response probability,
\[
\pi_\lambda(x,t,y)
=
\bP_\lambda(R^Y=1\mid X=x,T=t,Y=y,R^X=1,R^T=1)
=
\expit\{\lambda^\top Z(x,t,y)\},
\]
where the design vector $Z(x,t,y)$ is chosen to enforce the relevant missingness restriction:
under Assumption~\ref{ass2}, the vector $Z$ depends only on $(x,y)$;
and under Assumption~\ref{ass3}, $Z$ depends only on $(t,y)$.

For a unit with $R^Y=1$, its contribution to the likelihood is
\[
\bP_\beta(Y\mid X,T,R^X=1,R^T=1)\,
\pi_\lambda(X,T,Y).
\]
For a unit with $R^Y=0$, the outcome is latent and the contribution is
\[
\int \bP_\beta(y\mid X,T,R^X=1,R^T=1)\,
\{1-\pi_\lambda(X,T,y)\}\,dy.
\]
We employ the EM algorithm to handle the latent outcome, in the presence of outcome missingness.

In the E-step, for a unit with $R^Y=0$, the conditional distribution of the missing outcome under the current parameters is
\[
\bP_{\beta,\lambda}(y\mid X,T,R^Y=0,R^X=1,R^T=1)
\ \propto\
\bP_\beta(y\mid X,T,R^X=1,R^T=1)\,
\{1-\pi_\lambda(X,T,y)\}.
\]
If $Y$ is discrete, this conditional distribution can be computed exactly.
When $Y$ is continuous, direct evaluation is typically infeasible.
We therefore focus on fractional imputation (FI) \citep{kim2011parametric},
which provides a convenient approximation to the E-step of the EM algorithm in this setting.

\begin{itemize}[leftmargin=1.5em]
\item \emph{I-step (imputation).}
For each unit $i$ with $R_i^Y=0$, draw $M$ candidate values
$y_{ij}^\ast\sim h(\cdot\mid X_i,T_i)$, $j=1,\ldots,M$, from a proposal distribution $h$. A convenient default is to take $h$ as the complete-case outcome model,
\[
h(\cdot\mid X_i,T_i)=\bP_{\beta^{(0)}}(\cdot\mid X_i,T_i,R_i^X=1,R_i^T=1),
\]
where $\beta^{(0)}$ is obtained by fitting $\bP_\beta(\cdot\mid X,T,R^X=1,R^T=1)$ on units with
$R^Y=1$. Further discussion of the choice of the proposal distribution $h$ is deferred to
\citet{yang2016fractional}.


\item \emph{W-step (weighting).}
Compute fractional weights
\[
w_{ij}^{(m)} \propto 
\frac{\bP_{\beta^{(m)}}(y_{ij}^\ast\mid X_i,T_i,R_i^X=1,R_i^T=1)\,
\{1-\pi_{\lambda^{(m)}}(X_i,T_i,y_{ij}^\ast)\}}
     {h(y_{ij}^\ast\mid X_i,T_i)},
\qquad \sum_{j=1}^M w_{ij}^{(m)}=1.
\]
\item \emph{M-step (maximization).}
Maximize the Monte Carlo approximation to the EM $Q$-function:
\begin{align*}
Q^{(m)}(\beta,\lambda)
&=
\sum_{i:\,R_i^Y=1}
\bigl\{\log\bP_\beta(Y_i\mid X_i,T_i,R_i^X=1,R_i^T=1)
+\log\pi_\lambda(X_i,T_i,Y_i)\bigr\} +\\
&\quad
\sum_{i:\,R_i^Y=0}\sum_{j=1}^M w_{ij}^{(m)}
\bigl\{\log\bP_\beta(y_{ij}^\ast\mid X_i,T_i,R_i^X=1,R_i^T=1)
+\log[1-\pi_\lambda(X_i,T_i,y_{ij}^\ast)]\bigr\}.
\end{align*}
\end{itemize} 
We iterate until convergence. After obtaining $\widehat{\beta}$, we estimate $\tau_{t_1,t_0}(x)$ accordingly.
We again employ the bootstrap for inference.

\section{Simulation study}\label{sec:sim}
This section evaluates how different estimators recover $\tau_{t_1,t_0}(x)$ under Assumptions~\ref{ass1}--\ref{ass3}.

\subsection{Setups and estimators}
Each simulation uses $N=1000$ observations and is repeated 500 times. We consider eight data-type combinations, corresponding to all binary/continuous choices of $X$, $T$, and $Y$. Performance is summarized by the percent bias of $\tau_{t_1,t_0}(x)$ at a fixed $x$ with $(t_1,t_0)=(1,0)$. Parameters are calibrated so that the marginal missingness rates of $X$, $T$, and $Y$ are each approximately 20\%. Full details are provided in Section~\ref{app:sim-more} of the supplementary material.

We compare the following estimators:
\begin{itemize}
\item \emph{Oracle}: fits the correctly specified outcome model $\bP_\beta(Y\mid T,X)$ to the full data by plugging in the true values of the missing data.

\item \emph{CCA}: fits the correctly specified outcome model $\bP_\beta(Y\mid T,X)$ using only complete cases, i.e., units with $R^X = R^T = R^Y = 1$.

\item \emph{$X$-miss-indicator + CCA}: augments the covariate with the missingness indicator $\mathbb{I}(R^X=0)$, restricts the analysis to units with observed treatment and outcome $(R^T=R^Y=1)$, and fits the correctly specified outcome model for $Y$ given $T$ and the augmented covariates.

\item \emph{MI (restricted)}: a MI procedure \citep{vanbuuren2011multivariate} with predictive imputation models that exclude $Y$ when imputing $X$ and $T$, ensuring the outcome does not drive reconstruction of the covariate and treatment, consistent with the design-stage principle of \citet{Rubin2007}. The outcome $Y$ is imputed using the same model as in the outcome analysis.

\item \emph{MI (all)}: a MI procedure \citep{vanbuuren2011multivariate} that conditions on all other variables when imputing each of $X$, $T$, and $Y$.

\item \emph{NP}: the proposed nonparametric estimator.

\item \emph{Para}: the proposed parametric estimator.

\item \emph{Para (full)}: unlike the subset analysis used in \emph{Para}, this approach additionally specifies parametric models for $(X,T,R^X,R^T)$ and maximizes the full observed-data likelihood.  This is included as a cautionary benchmark, since the full joint distribution is generally not identified under Assumptions~\ref{ass1}--\ref{ass3}.

\end{itemize}

\subsection{Results}
For each estimator, we obtain estimates $\widehat{\tau}_{t_1,t_0}(x)$ across $500$ simulation replicates. We summarize performance using replicate-level percent bias
\[
\text{Percent bias}^{(r)}
=
100\times\frac{\widehat{\tau}^{(r)}_{t_1,t_0}(x)-\tau_{t_1,t_0}(x)}{\tau_{t_1,t_0}(x)},
\qquad r=1,\dots,500.
\]
We report the distribution of this quantity using boxplots. The
center of each box reflects the mean percent bias across replicates, and the
spread reflects the variability of the estimator. We also examine a null effect setting with $\tau_{t_1,t_0}(x)=0$, corresponding to part (ii) of Theorem~\ref{the2}. Since percent bias is not well-defined when $\tau_{t_1,t_0}(x)=0$, we instead assess performance using the difference $\widehat{\tau}_{t_1,t_0}(x)-\tau_{t_1,t_0}(x)$. Results for this null effect setting are reported in Section~\ref{app:sim-more} of the supplementary material and are consistent with the theory.

\begin{figure}[H]
    \centering
    \includegraphics[width=\linewidth]{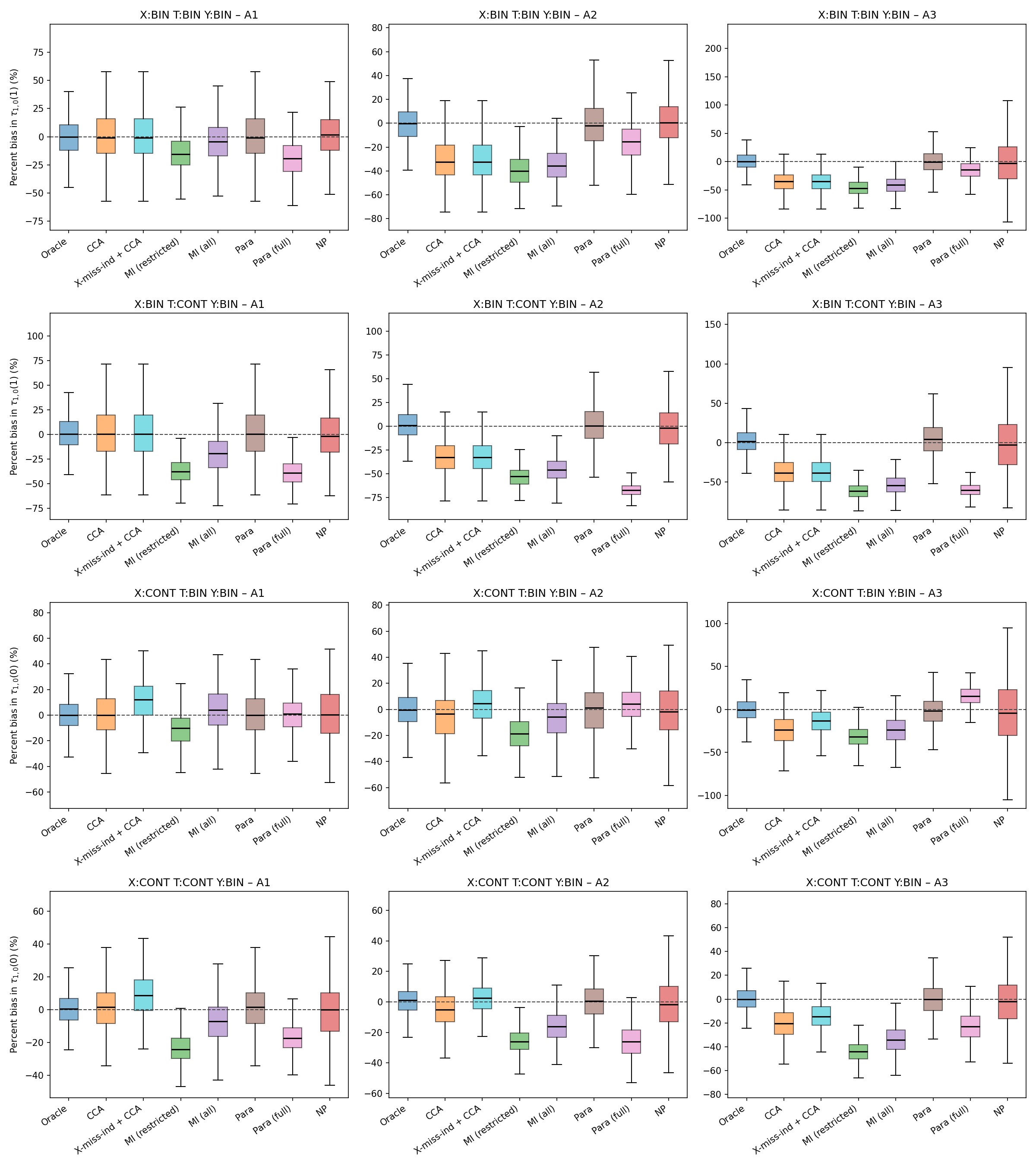}
\caption{CCA is unbiased only under Assumption~\ref{ass1}; under Assumptions~\ref{ass2}--\ref{ass3}, bias can be substantial. Boxplots show percent bias in $\tau_{1,0}(1)$ (binary $X$) and $\tau_{1,0}(0)$ (continuous $X$) with binary $Y$; closer to zero is better. Rows correspond to $(X,T)$ type combinations and columns to Assumptions~\ref{ass1}--\ref{ass3}. Methods: Oracle, CCA, $X$-miss-indicator + CCA, MI (restricted), MI (all), NP, Para, and Para (full).}

    \label{fig:sim1}
\end{figure}

\begin{figure}[H]
    \centering
    \includegraphics[width=\linewidth]{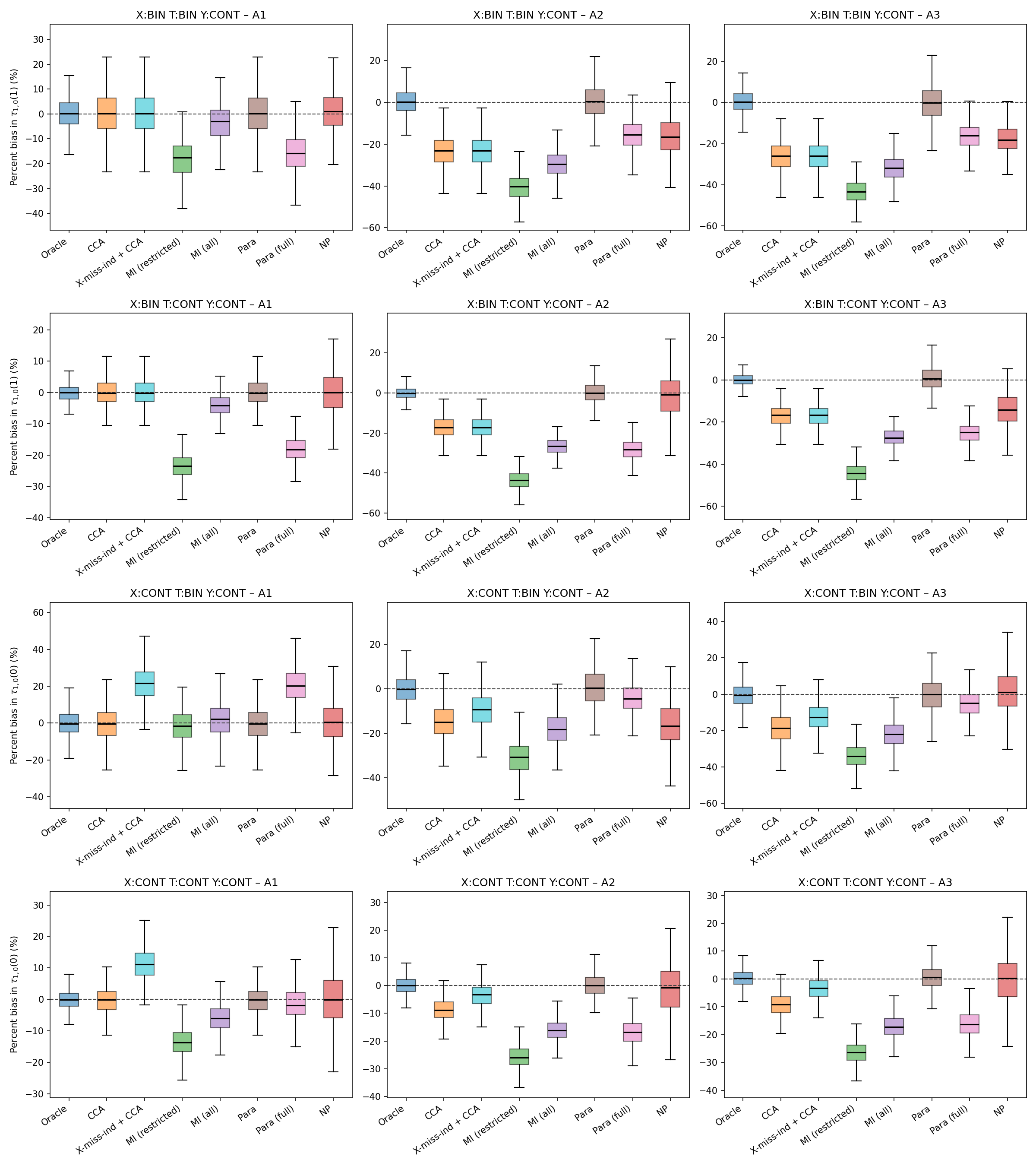}
\caption{Under Assumptions~\ref{ass2}--\ref{ass3}, estimators that ignore outcome-dependent missingness can be biased. Boxplots show percent bias in $\tau_{1,0}(1)$ (binary $X$) and $\tau_{1,0}(0)$ (continuous $X$) with continuous $Y$; closer to zero is better. Rows correspond to $(X,T)$ type combinations and columns to Assumptions~\ref{ass1}--\ref{ass3}. Methods: Oracle, CCA, $X$-miss-indicator + CCA, MI (restricted), MI (all), NP, Para, and Para (full).}

    \label{fig:sim2}
\end{figure}

For settings with $\tau_{t_1,t_0}(x)\neq 0$, percent bias is shown in Figure~\ref{fig:sim1} for binary $Y$ and in Figure~\ref{fig:sim2} for continuous $Y$. Rows correspond to $(X,T)$ data type combinations and columns to Assumptions~\ref{ass1}--\ref{ass3}.

Under Assumption~\ref{ass1}, $\bP(Y\mid X,T)$ is identifiable from the complete cases, so CCA, NP, and Para remain nearly unbiased across all eight data-type combinations, while other estimators exhibit noticeable bias in some settings.

Under Assumptions~\ref{ass2} and~\ref{ass3}, identification of $\bP(Y\mid X,T)$ additionally requires the corresponding completeness condition. In Figure~\ref{fig:sim1}, and in Figure~\ref{fig:sim2} for the cases with continuous $T$ under
Assumption~\ref{ass2} and continuous $X$ under Assumption~\ref{ass3}, the completeness condition holds, and both NP and Para stay nearly unbiased. In contrast, in Figure~\ref{fig:sim2} for the cases with binary $T$ under Assumption~\ref{ass2}
and binary $X$ under Assumption~\ref{ass3}, the completeness condition fails without further assumptions. In these settings, the Fredholm equation underlying NP is underdetermined in the nonparametric model and NP can exhibit nontrivial bias, consistent with the theory. In contrast, Para remains unbiased because the completeness condition holds under the parametric models employed (see the parametric completeness examples in Section~\ref{app:param-ex} of the supplementary material). Other estimators show nontrivial biases. In general, NP displays larger variability than Para, as expected from a fully nonparametric approach that estimates and inverts an ill-posed integral equation. 

The performance of Para (full) illustrates that modeling additional components of the joint distribution, which are not generally identifiable, can lead to biased estimates. This reinforces the importance of subset analyses that directly target the identified object $\bP(Y\mid X,T)$ in Para and NP.


\section{Application to the NJCS and sensitivity analysis}\label{sec:sensitivity}

\subsection{Data and discussion on assumptions}
The data describe $8,707$ eligible applicants in the mid-1990s who lived in areas selected for in-person baseline interviews. Subjects were randomized to Job Corps program or control groups in the NJCS \citep{schochet2001national}. While the original study focused on the effect of the Job Corps program, here we seek to understand the causal relationship between post-baseline educational attainment and subsequent earnings.

Let $X$ denote baseline covariates, including program assignment, gender, age, race, education, prior-year earnings, parenthood, and ever arrested; in this application, all components of $X$ are coded as categorical indicators. Let $T$ be a binary indicator for obtaining an education or vocational credential at 30 months of follow-up, and let $Y$ denote weekly earnings at 48 months of follow-up. Our goal is to estimate how the effect of obtaining a credential $(T)$ on weekly earnings $(Y)$ varies with baseline covariates $X$, that is, the CATE $\tau_{t_1,t_0}(x)$ with $t_1=1$ and $t_0=0$.

This setting aligns with our prospective structure. The baseline covariates $X$ and $T$ (measured at 30 months) are both measured before $Y$ (48 months), so $Y$ cannot influence missingness in $X$ or $T$. These data exhibit substantial multivariate missingness. Several baseline covariates have missing values (education $\approx 0.65\%$, prior-year earnings $\approx 9.6\%$, ever arrested $\approx 6.7\%$), and approximately $21\%$ of $T$ and $21\%$ of $Y$ are missing. Consequently, appropriately addressing missingness in $(X,T,Y)$ is central to the analysis. 

We now discuss the plausibility of each assumption. Assumption \ref{ass1} requires that earnings do not affect their own missingness. This assumption may be questionable in practice, since earnings information is often sensitive. In survey settings, respondents may selectively withhold earnings depending on the earnings values themselves. Assumption \ref{ass2} requires that educational attainment does not directly affect earnings missingness except through missingness in other variables or earnings themselves. This assumption is arguably reasonable because respondents' willingness to disclose earnings likely depends on their general attitudes toward sensitive information (reflected in the missingness across other variables) or the earnings values themselves. The completeness condition in Theorem \ref{the2} further requires that educational attainment influences earnings. Empirical support for this assumption comes from prior works. \citet{zuo2025mediation} and \citet{qin2019multisite, qin2021unpacking} documented indirect effects of Job Corps on earnings through educational attainment, providing evidence of a nonzero treatment effect in our setting. However, previous analyses employed more restrictive assumptions regarding missing data. We instead focus on the CATE of educational attainment under weaker and more flexible assumptions.

Assumption \ref{ass3} in its multivariate extension (Remark~\ref{rem:idX}) requires the existence of baseline covariates that do not directly affect outcome missingness, except through covariate missingness, educational attainment, its missingness, or earnings. As an illustration, we take race and decompose $X=(X^{\mathrm{id}},X^c)$, where $X^{\mathrm{id}}$ denotes the race variable and $X^c$ collects the remaining baseline covariates. The associated completeness condition in Theorem \ref{the3} further requires that race is associated with earnings.

Because these missingness mechanisms are intrinsically untestable, it is essential to assess the robustness of causal conclusions to different missingness assumptions. In addition, recognizing that all three missingness mechanisms may be violated in practice, we conduct sensitivity analyses to evaluate the stability of our findings under quantified departures from each assumption.

\subsection{Models}
We apply both nonparametric and parametric approaches. Because weekly earnings contain a substantial point mass at zero (14.85\%), we adopt a two-part model that separately estimates the probability of positive earnings and the magnitude of positive earnings. Let $D=\mathbb{I}(Y>0)$ indicate whether an individual has any earnings. Under Assumptions \ref{ass2} and \ref{ass3}, missingness can depend on $Y$. We assume this dependence operates through the binary indicator $D$ (zero versus positive earnings) rather than through the magnitude of positive earnings conditional on $D=1$. This restriction allows the completeness conditions in Theorems~\ref{the2} and~\ref{the3}  to hold with discrete support for $T$ and $X^{\mathrm{id}}$ even without additional parametric assumptions, provided $D$ fully captures the information relevant to missingness.

We model the probability for positive earnings using logistic regression:
\[
\logit\,\bP(D=1\mid T,X)
= \alpha + \beta_T\,T + W(X)^\top\gamma + \{T\cdot W(X)\}^\top\eta,
\]
where $W(X)$ is a vector of dummy variables for the covariates (with one baseline category per covariate), and $T\cdot W(X)$ denotes treatment-covariate interactions.
Denote $p(t,x)=\bP(D=1\mid T{=}t,X{=}x).$

We model the conditional mean of $Y$ among the positive earners ($D=1$) using Gamma regression with a log link:
\[
\log \bE\{Y\mid D=1,\,T,X\}
= \alpha_Y + \theta_T\,T + W(X)^\top\theta + \{T\cdot W(X)\}^\top\kappa.
\]
Since we assume the missingness depends on $Y$ only through $D$, conditioning on $D=1$ removes the need for further missingness correction. Consequently, the parametric estimator is fit using complete cases with $Y>0$, while the nonparametric estimator replaces the parametric functional form with nonparametric regression. Writing the conditional mean as $m(t,x)=\bE(Y\mid D=1,\,T{=}t,X{=}x),$ we have
$$\bE(Y\mid T{=}t,X{=}x)=p(t,x)\,m(t,x),$$
and therefore the CATE is
$$\tau_{t_1,t_0}(x)=p(t_1,x)\,m(t_1,x)-p(t_0,x)\,m(t_0,x).$$
To address outcome missingness, we specify a logistic model for the response indicator:
\[
\pi_\lambda(x,t,d)
= \bP_\lambda(R^Y=1\mid X=x,T=t,D=d,R^X=1,R^T=1)
= \expit\{\lambda^\top Z(x,t,d)\},
\]
where $Z(x,t,d)$ is chosen to match Assumptions~\ref{ass1}--\ref{ass3}. Under Assumption~\ref{ass3} with multivariate $X=(X^{\mathrm{id}},X^c)$, we enforce the weaker restriction \[R^Y \perp\!\!\!\perp X^{\mathrm{id}} \mid (T,D,R^X=1,R^T=1,X^c),\] so the baseline $R^Y$ model may depend on $X^c$ but excludes $X^{\mathrm{id}}$ (race).

We report $\tau_{t_1,t_0}(x)$ evaluated at a reference profile $x_{\text{ref}}$,
chosen as the most common category of each baseline covariate among units with observed treatment and covariates: assigned to Job Corps program, male gender, age between 16 and 17, black race, no child, never arrested, no high
school diploma or GED, and zero prior earnings.
The target estimand is thus $\tau_{1,0}(x_\text{ref})$.

We compare six estimation approaches: CCA, $X$-miss-indicator + CCA, MI (restricted), MI (all), NP, and Para. 
For each estimator, we compute $\widehat\tau_{1,0}(x_{\mathrm{ref}})$ with 95\% bootstrap percentile confidence intervals (CIs) based on 500 bootstrap resamples.

\subsection{Results.}
\begin{figure}[H]
  \centering
  \includegraphics[width=1\linewidth]{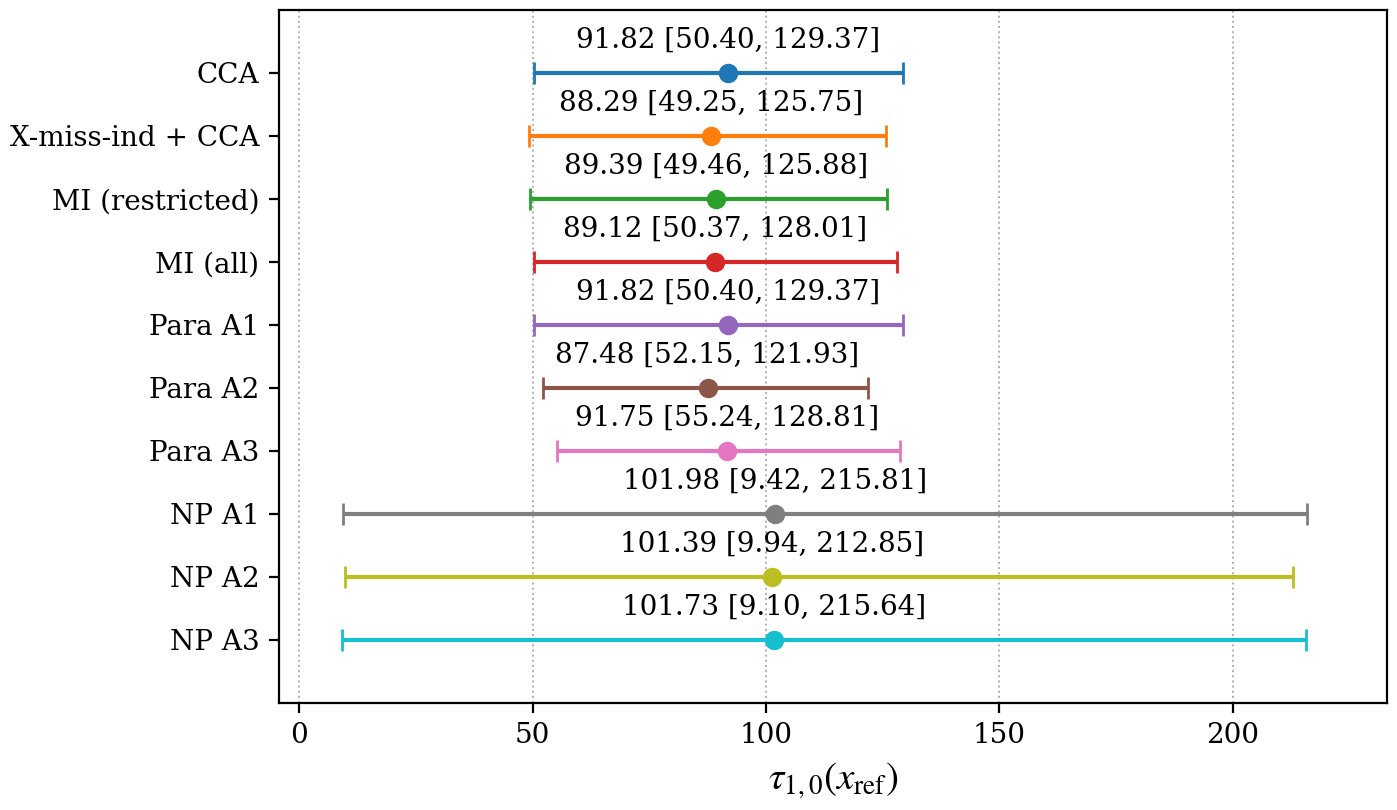}
\caption{Credential attainment increases earnings for the reference profile: all 95\% bootstrap percentile CIs for $\tau_{1,0}(x_\text{ref})$ exclude zero. Shown are CCA, $X$-miss-indicator + CCA, MI (restricted), MI (all), NP, and Para; NP and Para are reported under Assumptions~\ref{ass1}--\ref{ass3}. Larger values indicate larger earnings gains.}
\label{fig:njcs-earnings-cate-base}
\end{figure}

Figure~\ref{fig:njcs-earnings-cate-base} summarizes treatment effect estimates for the reference group. We report results for six estimators (CCA, X-miss-ind + CCA, MI(restricted), MI(all), NP, and Para), with NP and Para estimates shown separately under Assumptions~\ref{ass1}, \ref{ass2}, and \ref{ass3}. Despite this methodological and assumption heterogeneity, all specifications yield the same qualitative conclusion: the 95\% bootstrap CIs for $\tau_{1,0}(x_{\mathrm{ref}})$ exclude zero, confirming a positive earnings effect. Point estimates vary modestly across estimators, while the nonparametric approach produces wider CIs due to its greater estimation variability.

\subsection{Sensitivity analysis}

\begin{figure}[t]
  \centering
  \begin{subfigure}[t]{0.32\linewidth}
    \centering
    \begin{tikzpicture}[>=latex, scale=0.9, every node/.style={scale=0.9}]
      \node (x) at (0,0) {$X$}; \node (t) at (1.7,0) {$T$}; \node (y) at (3.4,0) {$D$};
      \node (rx) at (0,1.8) {$R^X$}; \node (rt) at (1.7,1.8) {$R^T$}; \node (ry) at (3.4,1.8) {$R^Y$};
      \path[-latex] (x) edge (t) (x) edge[bend right] (y) (t) edge (y);
      \path[-latex] (x) edge (rx) (x) edge (rt) (t) edge (rx) (t) edge (rt) (rx) edge (rt);
      \path[-latex] (x) edge (ry) (t) edge (ry) (rt) edge (ry) (rx) edge[bend left] (ry);
      \path[-latex, red, dotted, thick] (y) edge (ry); 
    \end{tikzpicture}
\caption*{Assumption 1 departure: allow $D\!\to\!R^Y$ (offset $\delta\,D$).}

  \end{subfigure}\hfill
  \begin{subfigure}[t]{0.32\linewidth}
    \centering
    \begin{tikzpicture}[>=latex, scale=0.9, every node/.style={scale=0.9}]
      \node (x) at (0,0) {$X$}; \node (t) at (1.7,0) {$T$}; \node (y) at (3.4,0) {$D$};
      \node (rx) at (0,1.8) {$R^X$}; \node (rt) at (1.7,1.8) {$R^T$}; \node (ry) at (3.4,1.8) {$R^Y$};
      \path[-latex] (x) edge (t) (x) edge[bend right] (y) (t) edge (y);
      \path[-latex] (x) edge (rx) (x) edge (rt) (t) edge (rx) (t) edge (rt) (rx) edge (rt);
      \path[-latex] (x) edge (ry) (y) edge (ry) (rt) edge (ry) (rx) edge[bend left] (ry);
      \path[-latex, red, dotted, thick] (t) edge (ry); 
    \end{tikzpicture}
    \caption*{Assumption 2 departure: allow $T\!\to\!R^Y$ (offset $\delta\,T$).}
  \end{subfigure}\hfill
  \begin{subfigure}[t]{0.32\linewidth}
    \centering
    \begin{tikzpicture}[>=latex, scale=0.9, every node/.style={scale=0.9}]
      \node (x) at (0,0) {$X_{\mathrm{id}}$}; \node (t) at (1.7,0) {$T$}; \node (y) at (3.4,0) {$D$};
      \node (rx) at (0,1.8) {$R^X$}; \node (rt) at (1.7,1.8) {$R^T$}; \node (ry) at (3.4,1.8) {$R^Y$};
      \path[-latex] (x) edge (t) (x) edge[bend right] (y) (t) edge (y);
      \path[-latex] (x) edge (rx) (x) edge (rt) (t) edge (rx) (t) edge (rt) (rx) edge (rt);
      \path[-latex] (t) edge (ry) (y) edge (ry) (rt) edge (ry) (rx) edge[bend left] (ry);
      \path[-latex, red, dotted, thick] (x) edge (ry); 
    \end{tikzpicture}
    \caption*{Assumption 3 departure: allow $X^{\mathrm{id}}\!\to\!R^Y$ (offset $\delta\,s(X^{\mathrm{id}})$).}
  \end{subfigure}
\caption{Sensitivity analysis adds one excluded edge into $R^Y$ to quantify departures from each baseline MNAR assumption. Each dotted red arrow indicates the violation introduced by offset~$\delta$. At $\delta=0$, the baseline missingness model holds. In the Assumption~\ref{ass3} panel, $X^{\mathrm{id}}$ denotes race (the identifying covariate component); the remaining baseline covariates $X^c$, which may influence any other variables in the DAG, are omitted for readability.}
  \label{fig:njcs-sens-dags}
\end{figure}

Assumptions~\ref{ass1}--\ref{ass3} impose restrictions on how $R^Y$ may depend on $(X,T,D)$, but these restrictions are not testable from the observed data. When an additional dependence is introduced (as in Figure~\ref{fig:njcs-sens-dags}), the observable data no longer identify $\tau_{t_1,t_0}(x)$ without further assumptions. To assess robustness, a sensitivity analysis is conducted using the parametric approach by embedding each missingness model in a single-parameter family indexed by an offset $\delta$ added to the logit for $R^Y$. Denote the baseline logit by $g(\cdot)$. We consider
\[
\begin{aligned}
\textup{Assumption 1:}~ 
&\logit\,\bP(R^Y{=}1\mid X,T,D,R^X=1,R^T=1)
   = g(X,T,R^X=1,R^T=1)+\delta\,D,\\[3pt]
\textup{Assumption 2:}~ 
&\logit\,\bP(R^Y{=}1\mid X,T,D,R^X=1,R^T=1)
   = g(X,D,R^X=1,R^T=1)+\delta\,T,\\[3pt]
\textup{Assumption 3:}~ 
&\logit\,\bP(R^Y{=}1\mid X,T,D,R^X=1,R^T=1)
   = g(T,D,X_c,R^X=1,R^T=1)+\delta\,s(X^{\mathrm{id}}),
\end{aligned}
\]
where $s(X^{\mathrm{id}})$ denotes the vector of dummies for the race variable.
At $\delta=0$ the corresponding missingness assumption holds. When $\delta\neq 0$, one additional directed edge ($D\!\to\!R^Y$, $T\!\to\!R^Y$, or $X_{\mathrm{id}}\!\to\!R^Y$) is introduced into the corresponding missingness mechanism, as illustrated in Figure~\ref{fig:njcs-sens-dags}.

We consider a broad range of sensitivity values, with $\delta \in [-2,2]$ on the log-odds scale \citep{chen2010big}. For each sensitivity value, we refit the proposed parametric models under the corresponding offset value to obtain $\widehat p(t,x_{\mathrm{ref}};\delta)$. We then report the implied effect
$$\widehat\tau_{1,0}(x_\text{ref};\delta)
=\widehat p(1,x_\text{ref};\delta)\,\widehat m(1,x_\text{ref})
-\widehat p(0,x_\text{ref};\delta)\,\widehat m(0,x_\text{ref}).$$

\begin{figure}[t]
  \centering
  \includegraphics[width=1\linewidth]{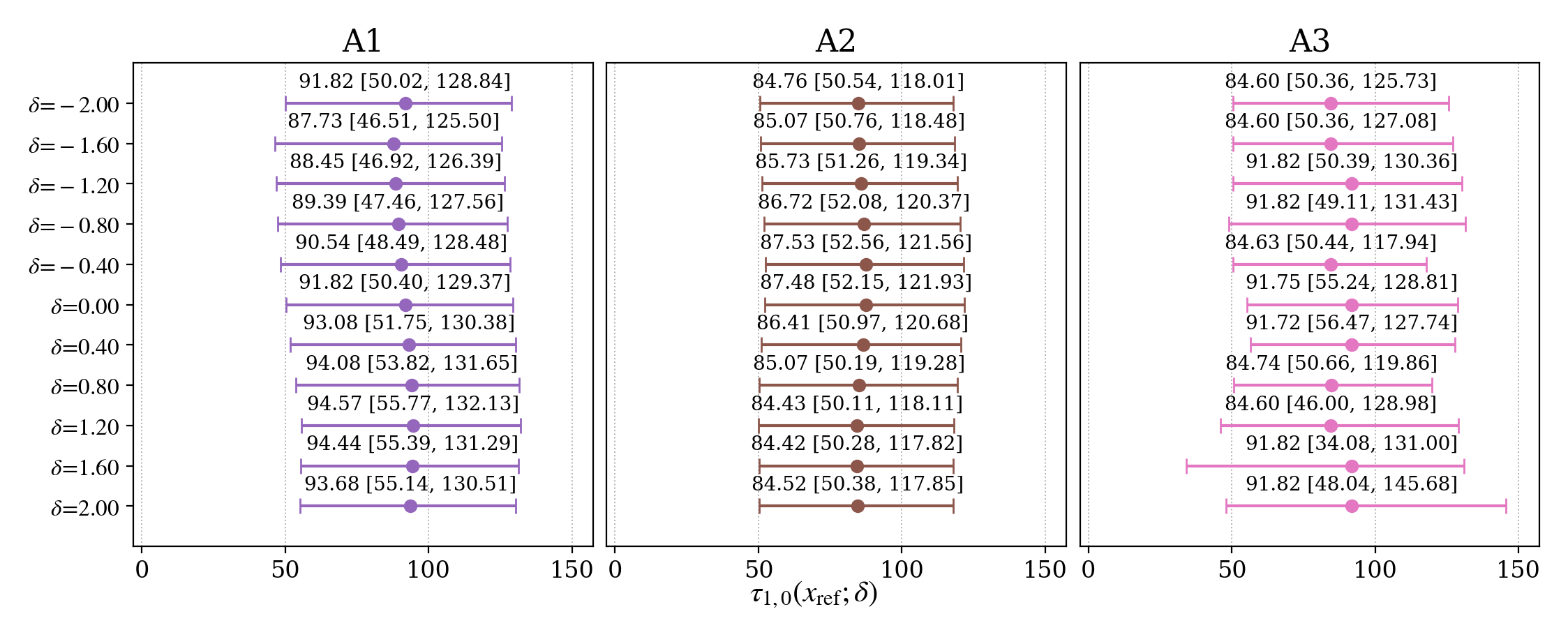}
\caption{The estimated effect remains positive across $\delta\in[-2,2]$ under all three sensitivity models. Shown are Para estimates of $\tau_{1,0}(x_\text{ref};\delta)$ with 95\% bootstrap percentile CIs under Assumptions~\ref{ass1}--\ref{ass3} and their single-parameter sensitivity variants indexed by $\delta\in[-2,2]$.}
\label{fig:njcs-earnings-cate-sen}
\end{figure}

Figure~\ref{fig:njcs-earnings-cate-sen} reports the sensitivity analysis indexed by $\delta\in[-2,2]$. The estimated effect remains positive throughout this range, demonstrating substantial robustness to departures from each baseline assumption.

\section{Discussion}\label{sec:discussion}

This paper studies identification and estimation of the CATE in observational studies where covariates, treatment, and outcome may be MNAR. In prospective settings, the framework allows general missingness in covariates and treatment and considers three outcome-missingness mechanisms (including self-censoring) under which the CATE remains identifiable. We develop two estimation strategies. The nonparametric approach implements the identification results by solving an integral equation, but can be numerically unstable, sensitive to tuning choices, and less suited for sensitivity analysis. The parametric approach trades flexibility for stability via structured outcome and missingness models and admits a simple and interpretable sensitivity analysis.

In practice, we recommend using the parametric estimator as the primary analysis, reporting results under multiple missingness assumptions, and conducting sensitivity analysis. The nonparametric estimator then serves as a robustness check on the parametric specification rather than the default approach in applied analyses.

\section*{Data and code availability}
Data and code to reproduce all analyses are available at
\href{https://github.com/sushi133/CATE-MNAR}{\nolinkurl{https://github.com/sushi133/CATE-MNAR}}.

\section*{Acknowledgements} Shuozhi Zuo and Yixin Wang were supported in part by funding from the Office of Naval Research under grant N00014-23-1-2590, the National Science Foundation under grant No. 2310831, No. 2428059, No. 2435696, No. 2440954, a Michigan Institute for Data Science Propelling Original Data Science (PODS) grant, Two Sigma Investments LP, and  LG Management Development Institute AI Research. Any opinions, findings, and conclusions or recommendations expressed in this material are those of the authors and do not necessarily reflect the views of the sponsors.

\bibliographystyle{apalike}
\bibliography{ref}

\newpage
\bigskip
\begin{center}
{\bfseries Supplementary material}\\
\end{center}
\if0\blind
{}\fi

Section \ref{app:counterexample-general} presents two counterexamples referenced in the main paper.

Section \ref{sec::proofs} contains the proofs of all theorems.

Section \ref{app:param-ex} presents parametric examples for the completeness conditions.

Section \ref{app:np2sls-impl} details the nonparametric estimator.

Section \ref{app:sim-more} provides additional details and results on the simulation study.

\setcounter{page}{1}
\setcounter{equation}{0}
\setcounter{section}{0}
\setcounter{figure}{0}
\setcounter{table}{0}
\renewcommand {\thepage} {S\arabic{page}}
\renewcommand {\theequation} {S\arabic{equation}}
\renewcommand {\thesection} {S\arabic{section}}
\renewcommand{\thefigure}{S\arabic{figure}}
\renewcommand{\thetable}{S\arabic{table}}

\section{Counterexamples}\label{app:counterexample-general}

This section presents two counterexamples referenced in the main paper. The first establishes non-identifiability of the ATE under Assumptions~\ref{ass1}--\ref{ass3}. The second establishes non-identifiability of the CATE when $R^Y$ admits an additional dependence path beyond Assumptions~\ref{ass1}--\ref{ass3}.

\subsection{Counterexample 1}\label{app:counterexample-general1}

We allow general missingness in $(X,T)$ under Assumptions~\ref{ass1}--\ref{ass3} and assume that $Y$ is fully observed:
\[
P(X,T,Y,R^X,R^T)=P(X)\,P(T\mid X)\,P(Y\mid X,T)\,P(R^X\mid X,T)\,P(R^T\mid X,T,R^X).
\]
Assume $X,T,Y\in\{0,1\}$. For simplicity, consider the special case $R^T=R^X$,
so that $(X,T)$ are either jointly observed or jointly missing.

The observed data identify
\[
\begin{aligned}
p^{11}_{xt} &= P(X=x,T=t,Y=1,R^X=1,R^T=1),\\
p^{01}_{xt} &= P(X=x,T=t,Y=0,R^X=1,R^T=1),\\
p^{+1} &= P(Y=1,R^X=0,R^T=0),\qquad
p^{+0} = P(Y=0,R^X=0,R^T=0),
\end{aligned}
\]
with $\sum_{x,t}(p^{11}_{xt}+p^{01}_{xt})+p^{+1}+p^{+0}=1$.
In this counterexample,
\[
\begin{aligned}
&(p^{11}_{00},p^{01}_{00})
=\Bigl(\frac{45}{800},\,\frac{180}{800}\Bigr),\qquad
(p^{11}_{01},p^{01}_{01})
=\Bigl(\frac{72}{800},\,\frac{18}{800}\Bigr),\\
&(p^{11}_{10},p^{01}_{10})
=\Bigl(\frac{28}{800},\,\frac{7}{800}\Bigr),\qquad
(p^{11}_{11},p^{01}_{11})
=\Bigl(\frac{12}{800},\,\frac{48}{800}\Bigr),\\
&(p^{+1},p^{+0})
=\Bigl(\frac{123}{800},\,\frac{267}{800}\Bigr).
\end{aligned}
\]

Let $\gamma_{xt}=P(X=x,T=t)$, $\theta_{xt}=P(Y=1\mid X=x,T=t)$, and
$\pi_{xt}=P(R^X=1,R^T=1\mid X=x,T=t)$.
Then
\[
p^{11}_{xt}=\gamma_{xt}\,\theta_{xt}\,\pi_{xt},\qquad
p^{01}_{xt}=\gamma_{xt}\,(1-\theta_{xt})\,\pi_{xt}.
\]
Thus the observed data identify $\theta_{xt}=p^{11}_{xt}/(p^{11}_{xt}+p^{01}_{xt})$
but do not identify $\gamma_{xt}$, since only the products $\gamma_{xt}\pi_{xt}$ are observed.

We now construct two different parameterizations, models A and B, that both
reproduce the same observed data probabilities but imply different ATEs. Both models
share the same conditional outcome distribution:
\[
(\theta_{00},\theta_{01},\theta_{10},\theta_{11})
=\Bigl(\frac15,\frac45,\frac45,\frac15\Bigr).
\]

Model A has
\[
P(X=1)=\frac14,\qquad P(T=1\mid X=0)=\frac14,\qquad P(T=1\mid X=1)=\frac34,
\]
with response probabilities
\[
(\pi_{00},\pi_{01},\pi_{10},\pi_{11})
=\Bigl(\frac12,\frac35,\frac{7}{10},\frac25\Bigr).
\]

Model B has
\[
P(X=1)=\frac12,\qquad P(T=1\mid X=0)=\frac{17}{50},\qquad P(T=1\mid X=1)=\frac{21}{25},
\]
with response probabilities
\[
(\pi_{00},\pi_{01},\pi_{10},\pi_{11})
=\Bigl(\frac{75}{88},\frac{45}{68},\frac{35}{64},\frac{5}{28}\Bigr).
\]

Both parameterizations yield the same observed data distribution
$(p^{11}_{xt},p^{01}_{xt},p^{+1},p^{+0})$.
However, with the shared $\theta_{xt}$ above, $\theta_{01}-\theta_{00}=\frac35$ and $\theta_{11}-\theta_{10}=-\frac35$, so
\[
\mathrm{ATE}=\frac35\{1-2P(X=1)\}.
\]
Therefore,
\[
\mathrm{ATE}_A=\frac{3}{10}\qquad\text{but}\qquad \mathrm{ATE}_B=0,
\]
showing that the ATE is not identifiable from the observed data under the missingness mechanisms for $(X,T)$ under Assumptions~\ref{ass1}--\ref{ass3}.

\subsection{Counterexample 2}\label{app:counterexample-general2}

We consider $X$ and $T$ fully observed, while allowing an additional dependence path into $R^Y$ beyond Assumptions~\ref{ass1}--\ref{ass3}:
\[
\bP(X,T,Y,R^Y)
= \bP(X)\,\bP(T\mid X)\,\bP(Y\mid X,T)\,\bP(R^Y\mid X,T,Y).
\]
Assume $X,T,Y\in\{0,1\}$.
The observed data identify
\[
\begin{aligned}
p^{11}_{xt} &= \bP(X=x,T=t,Y=1,R^Y=1),\\
p^{01}_{xt} &= \bP(X=x,T=t,Y=0,R^Y=1),\\
p^{+0}_{xt} &= \bP(X=x,T=t,R^Y=0),
\end{aligned}
\]
with $p^{11}_{xt}+p^{01}_{xt}+p^{+0}_{xt}=\bP(X=x,T=t)$.
In our counterexample, 
\[
\begin{aligned}
&(p^{11}_{00},p^{01}_{00},p^{+0}_{00})
=\Bigl(\tfrac{30}{800},\,\tfrac{192}{800},\,\tfrac{78}{800}\Bigr),\\
&(p^{11}_{01},p^{01}_{01},p^{+0}_{01})
=\Bigl(\tfrac{24}{800},\,\tfrac{28}{800},\,\tfrac{48}{800}\Bigr),\\
&(p^{11}_{10},p^{01}_{10},p^{+0}_{10})
=\Bigl(\tfrac{12}{800},\,\tfrac{36}{800},\,\tfrac{52}{800}\Bigr),\\
&(p^{11}_{11},p^{01}_{11},p^{+0}_{11})
=\Bigl(\tfrac{126}{800},\,\tfrac{81}{800},\,\tfrac{93}{800}\Bigr).
\end{aligned}
\]
Let $\theta_{xt}=\bP(Y=1\mid X=x,T=t)$ and
$\pi_{xty}=\bP(R^Y=1\mid X=x,T=t,Y=y)$.
Then, 
\begin{align*}
p^{11}_{xt} &= \bP(X=x,T=t)\,\theta_{xt}\,\pi_{xt1},\\
p^{01}_{xt} &= \bP(X=x,T=t)\,(1-\theta_{xt})\,\pi_{xt0}.
\end{align*}
Thus, the observed data identify only the products
$\theta_{xt}\pi_{xt1}$ and $(1-\theta_{xt})\pi_{xt0}$, leaving $\theta_{xt}$ unidentified without further assumptions.

We now construct two parameterizations, models A and B, that both reproduce the same observed data probabilities but imply different CATEs.
Both models share the same $(X,T)$ distribution:
\[
\bP(X=1)=\tfrac12,\qquad
\bP(T=1\mid X=0)=\tfrac14,\qquad
\bP(T=1\mid X=1)=\tfrac34.
\]

Model A has
\[
(\theta_{00},\theta_{01},\theta_{10},\theta_{11})
=\Bigl(\tfrac15,\tfrac35,\tfrac25,\tfrac7{10}\Bigr),
\]
with response probabilities
\[
(\pi_{000},\pi_{001},\pi_{010},\pi_{011},\pi_{100},\pi_{101},\pi_{110},\pi_{111})
=\Bigl(\tfrac45,\tfrac12,\tfrac7{10},\tfrac25,\tfrac35,\tfrac3{10},\tfrac9{10},\tfrac35\Bigr).
\]

Model B has
\[
(\theta_{00},\theta_{01},\theta_{10},\theta_{11})
=\Bigl(\tfrac3{10},\tfrac12,\tfrac12,\tfrac35\Bigr),
\]
with response probabilities
\[
(\pi_{000},\pi_{001},\pi_{010},\pi_{011},\pi_{100},\pi_{101},\pi_{110},\pi_{111})
=\Bigl(\tfrac{32}{35},\tfrac13,\tfrac{14}{25},\tfrac{12}{25},\tfrac{18}{25},\tfrac6{25},\tfrac{27}{40},\tfrac7{10}\Bigr).
\]

Both parameterizations yield the same observed data distribution.
However, with
\[
\tau_{1,0}(x)=\theta_{x1}-\theta_{x0},
\]
\[
\tau_{1,0;A}(0)=\tfrac25,\quad \tau_{1,0;A}(1)=\tfrac3{10},
\qquad
\tau_{1,0;B}(0)=\tfrac15,\quad \tau_{1,0;B}(1)=\tfrac1{10}.
\]
Therefore, the same observed data are compatible with distinct values of the CATE, showing that the CATE is not identifiable under the general outcome missingness mechanism.

\section{Proofs}\label{sec::proofs}
Throughout this section, we maintain the causal assumptions $(i)$ to $(iii)$, so identification of $\bP(Y\mid T,X)$ implies identification of $\tau_{t_1,t_0}(x)$.

\subsection{Proof of Theorem \ref{the1}}
Identification of $\bP(Y=y\mid X=x,T=t)$ follows from 
$$
\bP(Y=y\mid X=x,T=t)
=
\bP(Y=y\mid X=x,T=t,R^X=1,R^T=1,R^Y=1).
$$ 

\subsection{Proof of Theorem \ref{the2}}

\paragraph{Part (i).}
We focus on identifying $\bP(R^Y=1\mid X=x,Y=y,R^X=1,R^T=1)$. Define 
\begin{eqnarray*}
    \bP_{y1\mid tx11} &=& \bP(Y=y,R^Y=1\mid T=t,X=x,R^T=1,R^X=1), \\
    \bP_{+0\mid tx11} &=& \bP(R^Y=0\mid T=t,X=x,R^T=1,R^X=1), \\
    \zeta_{x}(y) &=&  \frac{\bP(R^Y=0\mid X=x,Y=y,R^X=1,R^T=1)}{\bP(R^Y=1\mid X=x,Y=y,R^X=1,R^T=1)}.
\end{eqnarray*}
Since
$$
\bP_{y1\mid tx11}
=
\bP(Y=y\mid T=t,X=x)\,
\bP(R^Y=1\mid X=x,Y=y,R^X=1,R^T=1),
$$
it follows that
\begin{align}
\bP_{+0\mid tx11}
&= \int_{y\in \mathcal{Y}} \bP(Y=y,R^Y=0\mid T=t,X=x,R^T=1,R^X=1)\,\textup{d}y \nonumber\\
&=\int_{y\in \mathcal{Y}} \bP(Y=y\mid T=t,X=x)\,
\bP(R^Y=0\mid X=x,Y=y,R^X=1,R^T=1)\,\textup{d}y \nonumber\\
&=\int_{y\in \mathcal{Y}}\bP_{y1\mid tx11}\,
\frac{\bP(R^Y=0\mid X=x,Y=y,R^X=1,R^T=1)}{\bP(R^Y=1\mid X=x,Y=y,R^X=1,R^T=1)}\,\textup{d}y \nonumber\\
&=\int_{y\in \mathcal{Y}}\bP_{y1\mid tx11}\,\zeta_{x}(y)\,\textup{d}y. \nonumber
\end{align}
The uniqueness of solutions $\zeta_{x}(y)$ requires that 
$\bP(Y,R^Y=1\mid T,X=x,R^T=1,R^X=1)$ is complete in $T$ for $x$. For discrete $T$ and discrete $Y$, the completeness assumption is equivalent to Rank $(\Theta_{x})=K$, where $\Theta_{x}$ is a $J \times K$ matrix with $\bP_{y1\mid tx11}$ as the $(t,y)$th element. For binary $Y$, the rank condition further reduces to $T \not\independent  Y\mid X$. For continuous $T$ and continuous $Y$, the dimension of $T$ must be no smaller than that of $Y$ in general, as required by the completeness assumption.

We can subsequently identify $\bP(R^Y=1\mid X=x,Y=y,R^X=1,R^T=1)$ once $\zeta_{x}(y)$ is identified. Then, the identification of $\bP(Y=y\mid X=x,T=t)$ follows from 
\begin{align}
\bP(Y=y\mid X=x,T=t)
&=\bP(Y=y\mid X=x,T=t,R^X=1,R^T=1) \nonumber\\
&= \frac{\bP(Y=y,R^Y=1\mid X=x,T=t,R^X=1,R^T=1)}{\bP(R^Y=1\mid X=x,Y=y,R^X=1,R^T=1)}.\nonumber
\end{align}

\paragraph{Part (ii).}
For any $t_1,t_0$, we have 
\begin{eqnarray*} &&\bP(Y=y\mid X=x,T=t_1)-\bP(Y=y\mid X=x,T=t_0) \\&=& \frac{\left[ \begin{aligned} &\bP(Y=y,R^Y=1\mid X=x,T=t_1,R^X=1,R^T=1)\\[-0.5em] &-\bP(Y=y,R^Y=1\mid X=x,T=t_0,R^X=1,R^T=1) \end{aligned}\right]} {\bP(R^Y=1\mid X=x,Y=y,R^X=1,R^T=1)}. \end{eqnarray*}
When $Y \independent T \mid X=x$, the numerator does not depend on $T$ and is therefore zero for all $y$. Consequently, the null effect $\tau_{t_1,t_0}(x)=0$ is identified.

\subsection{Proof of Theorem \ref{the3}}
We focus on identifying $\bP(R^Y=1\mid T=t,Y=y,R^X=1,R^T=1)$. Define
\begin{eqnarray*}
    \bP_{y1\mid tx11} &=& \bP(Y=y,R^Y=1\mid T=t,X=x,R^T=1,R^X=1),\\
    \bP_{+0\mid tx11} &=& \bP(R^Y=0\mid T=t,X=x,R^T=1,R^X=1),\\
    \zeta_t(y) &=& \frac{\bP(R^Y=0\mid T=t,Y=y,R^X=1,R^T=1)}
                       {\bP(R^Y=1\mid T=t,Y=y,R^X=1,R^T=1)}.
\end{eqnarray*}
Since
\[
\bP_{y1\mid tx11}
=
\bP(Y=y\mid T=t,X=x)\,
\bP(R^Y=1\mid T=t,Y=y,R^X=1,R^T=1),
\]
it follows that
\begin{align}
\bP_{+0\mid tx11}
&= \int_{y\in\mathcal{Y}} \bP(Y=y,R^Y=0\mid T=t,X=x,R^T=1,R^X=1)\,dy \nonumber\\
&= \int_{y\in\mathcal{Y}} \bP(Y=y\mid T=t,X=x)\,
\bP(R^Y=0\mid T=t,Y=y,R^X=1,R^T=1)\,dy \nonumber\\
&= \int_{y\in\mathcal{Y}} \bP_{y1\mid tx11}\,
\frac{\bP(R^Y=0\mid T=t,Y=y,R^X=1,R^T=1)}
     {\bP(R^Y=1\mid T=t,Y=y,R^X=1,R^T=1)}\,dy \nonumber\\
&= \int_{y\in\mathcal{Y}} \bP_{y1\mid tx11}\,\zeta_t(y)\,dy. \nonumber
\end{align}
for each $t\in\mathcal{T}$. The uniqueness of solutions $\zeta_{t}(y)$ requires that $\bP(Y,R^Y=1\mid T=t,X,R^T=1,R^X=1)$ is complete in $X$ for all $t$. For discrete $X$ and discrete $Y$, the completeness assumption is equivalent to Rank $(\Theta_{t})=K$, where $\Theta_{t}$ is a $L \times K$ matrix with $\bP_{y1\mid tx11}$ as the $(x,y)$th element. For binary $Y$, the rank condition further reduces to $X \not\independent Y\mid T$. For continuous $X$ and continuous $Y$, the dimension of $X$ must be no smaller than that of $Y$ in general, as required by the completeness assumption.

We can subsequently identify $\bP(R^Y=1\mid T=t,Y=y,R^X=1,R^T=1)$ once $\zeta_{t}(y)$ is identified. Then, the identification of $\bP(Y=y\mid X=x,T=t)$ follows from \begin{align} \bP(Y=y\mid X=x,T=t) &=\bP(Y=y\mid X=x,T=t,R^X=1,R^T=1) \nonumber\\ &= \frac{\bP(Y=y,R^Y=1\mid X=x,T=t,R^X=1,R^T=1)}{\bP(R^Y=1\mid T=t,Y=y,R^X=1,R^T=1)}.\nonumber \end{align}

\section{Parametric examples for completeness conditions}\label{app:param-ex}

This section presents parametric examples under which the completeness conditions in
Theorems~\ref{the2} and~\ref{the3} hold. Theorem~2.2 in
\citeSupp{suppnewey2003instrumental} establishes a general completeness result for
exponential-family data distributions, which we state here for reference.

\begin{result}\label{res:complete-expfam}
The distribution $\bP(A,B)=\psi(B)\,h(A)\exp\{\lambda(A)^\top \eta(B)\}$ is complete in $A$ if
(i) $\psi(B)>0$, (ii) the support of $\lambda(A)$ contains an open set, and (iii) the mapping
$B\mapsto \eta(B)$ is one-to-one.
\end{result}

For illustration, we present parametric outcome models that satisfy the corresponding
completeness assumptions for Theorems~\ref{the2} and~\ref{the3}.

\begin{proposition}[(example for Theorem~\ref{the2})]\label{prop:param-the2}
For continuous $Y$, under the linear model
\[
Y=\beta_0+\beta_t T+\beta_x^\top X+\beta_{tx}\,T\cdot X+\varepsilon,
\qquad \varepsilon\sim N(0,\sigma^2),
\]
if $\beta_t+\beta_{tx}\,x\neq 0$ for a given $x$, then 
the conditional joint distribution
\begin{eqnarray*}
&&\bP(T,Y,R^T=1,R^Y=1\mid X=x,R^X=1)
\\&=&
\bP(Y\mid T,X=x)\,\bP(T,R^T=1,R^Y=1\mid X=x,R^X=1)
\\&=&
\frac{1}{(2\pi\sigma^2)^{1/2}}
\exp\!\left\{
-\frac{\bigl(Y-\beta_0-\beta_t T-\beta_x^\top x-\beta_{tx}T\cdot x\bigr)^2}{2\sigma^2}
\right\}
\bP(T,R^T=1,R^Y=1\mid X=x,R^X=1)
\end{eqnarray*}
is complete in $T$.
\end{proposition}

Proposition~\ref{prop:param-the2} follows from Result~\ref{res:complete-expfam} with
$A=T$ and $B=Y$ by rewriting the joint distribution above in exponential-family form.
Specifically, one can take
\[
\lambda(T)=T,
\qquad
\eta_x(Y)=\sigma^{-2}\bigl(\beta_t+\beta_{tx}x\bigr)\,Y.
\]

\begin{proposition}[(example for Theorem~\ref{the3})]\label{prop:param-the3}
For continuous $Y$, under the linear model
\[
Y=\alpha_0+\alpha_t T+\alpha_x X+\alpha_{tx}\,T\cdot X+\varepsilon,
\qquad \varepsilon\sim N(0,\sigma^2),
\]
if $\alpha_x+\alpha_{tx}\,t\neq 0$ for a given $t$, then 
the conditional joint distribution
\begin{eqnarray*}
&&\bP(X,Y,R^X=1,R^Y=1\mid T=t,R^T=1)
\\&=&
\bP(Y\mid X,T=t)\,\bP(X,R^X=1,R^Y=1\mid T=t,R^T=1)
\\&=&
\frac{1}{(2\pi\sigma^2)^{1/2}}
\exp\!\left\{
-\frac{\bigl(Y-\alpha_0-\alpha_t t-\alpha_x X-\alpha_{tx}t\cdot X\bigr)^2}{2\sigma^2}
\right\}
\bP(X,R^X=1,R^Y=1\mid T=t,R^T=1)
\end{eqnarray*}
is complete in $X$.
\end{proposition}

Proposition~\ref{prop:param-the3} follows from Result~\ref{res:complete-expfam} with
$A=X$ and $B=Y$ by rewriting the joint distribution above in exponential-family form.
Specifically, one can take
\[
\lambda(X)=X,
\qquad
\eta_t(Y)=\sigma^{-2}\bigl(\alpha_x+\alpha_{tx}t\bigr)\,Y.
\]

\section{Implementation details for the 2SLS estimator}\label{app:np2sls-impl}

This section provides implementation details for the series 2SLS estimator under Assumption~\ref{ass2}.
We focus on recovering the response-odds function $\zeta$ from the Fredholm equation and then using
$\widehat\zeta$ to construct selection-corrected weights and estimate $\tau_{t_1,t_0}(x)$.

\paragraph{Integral equation.}
The response odds function is
\begin{align*}
\zeta(x,y)
&=
\frac{\bP(R^Y=0\mid X=x,Y=y,R^X=1,R^T=1)}{\bP(R^Y=1\mid X=x,Y=y,R^X=1,R^T=1)},
\qquad
\pi(x,y)=\{1+\zeta(x,y)\}^{-1},
\end{align*}
and it solves
\begin{eqnarray}\label{eq:S_np2sls_fredholm_XT}
&&\bP(R^Y=0\mid T=t,X=x,R^X=1,R^T=1)
\nonumber\\&=&
\int \bP(Y=y,R^Y=1\mid T=t,X=x,R^X=1,R^T=1)\,\zeta(x,y)\,dy.
\end{eqnarray}

\paragraph{Sieve approximation and sample analogue.}
When $Y$ is continuous, $\zeta(x,y)$ is infinite dimensional, so we approximate
\[
\zeta(x,y)\approx h(y)^\top\beta(x),
\]
where $h(y)=(h_1(y),\ldots,h_J(y))^\top$ is a chosen sieve basis in $y$. We use a Hermite-type envelope
basis $h_j(y)=\exp(-\tilde y^2)\tilde y^{\,j-1}$ with $\tilde y=(y-\bar y)/s_y$ and $j=1,\ldots,J$,
where $\bar y$ and $s_y$ are the sample mean and sample standard deviation computed from complete cases. 
When $Y$ is binary, we use a saturated two-point basis.

Define
\begin{align}
p_{0x}(t)
&= \bP(R^Y=0 \mid T=t, X=x, R^X=1, R^T=1), \label{eq:S_pt10_XT}\\
p_{1x}(t)
&= \bP(R^Y=1 \mid T=t, X=x, R^X=1, R^T=1), \label{eq:S_pt11_XT}\\
H_x(t)
&= \E\!\left\{h(Y)\mid T=t, R^Y=1, X=x, R^X=1, R^T=1\right\}. \label{eq:S_H_XT}
\end{align}
When $X$ is discrete, \eqref{eq:S_np2sls_fredholm_XT} reduces to
\begin{equation}\label{eq:S_discreteX_moment}
p_{0x}(t) \approx p_{1x}(t)\,H_x(t)^\top \beta(x).
\end{equation}
Replacing $p_{0x}(t)$, $p_{1x}(t)$, and $H_x(t)$ by nonparametric estimates and stacking over the
$t$-values used in the sample analogue yields a linear system
$\widehat b_x\approx \widehat M_x\,\beta(x)$, where $\widehat b_x$ stacks $\widehat p_{0x}(t)$ and
$\widehat M_x$ stacks $\{\widehat p_{1x}(t)\widehat H_x(t)^\top\}$ row-wise.

When $X$ is continuous, we use a tensor-product sieve
\[
\zeta(x,y)\approx \phi(x,y)^\top\theta,
\qquad
\phi(x,y)=g(x)\otimes h(y),
\]
where $g(x)=(g_1(x),\ldots,g_{J_x}(x))^\top$ is a sieve basis in $x$ (we use the same Hermite-type
envelope basis on the standardized $x$). Thus $\phi(x,y)\in\mathbb{R}^{J_xJ}$ and
$\theta\in\mathbb{R}^{J_xJ}$. When both $X$ and $Y$ are continuous, we optionally apply an affine
whitening transformation to $(X,Y)$ based on complete cases before evaluating the Hermite-envelope
basis to improve numerical conditioning. 

Define
\begin{align*}
p_{0}(t,x)&=\bP(R^Y=0 \mid T=t, X=x, R^X=1, R^T=1),
\\
p_{1}(t,x)&=\bP(R^Y=1 \mid T=t, X=x, R^X=1, R^T=1),\\
H(t,x)
&= \E\!\left\{h(Y)\mid T=t, R^Y=1, X=x, R^X=1, R^T=1\right\}. 
\end{align*}
Then equation \eqref{eq:S_np2sls_fredholm_XT} implies the moment condition
\begin{equation}\label{eq:S_continuousX_moment}
p_{0}(t,x)
\approx
p_{1}(t,x)\,
(g(x)\otimes H(t,x))^\top \theta,
\end{equation}
whose sample analogue yields a linear system $\widehat b\approx \widehat M\,\theta$ in stacked form.

\paragraph{Regularized solution.}
Because \eqref{eq:S_np2sls_fredholm_XT} is an ill-posed inverse problem, noise in the first-stage
estimates can be amplified when solving for $\zeta$. Given $\widehat b_x$ and $\widehat M_x$, we obtain
$\widehat\beta(x)$ by solving the regularized least-squares problem
\[
\widehat\beta(x)
=
\arg\min_{\beta}\ \|\widehat b_x-\widehat M_x\beta\|_2^2
\quad\text{subject to}\quad
\beta^\top\Lambda\beta \le B,
\]
and analogously obtain $\widehat\theta$ in the tensor-sieve case. Here, $\Lambda$ is a positive
definite penalty matrix imposing a compactness restriction on the sieve coefficients \citep{yang2019causal}.
The sieve dimensions and the bound $B$ jointly control the flexibility of $\widehat\zeta$:
larger $J$ (and $J_x$) can reduce approximation bias but often worsen conditioning and amplify sampling
noise, leading to unstable implied weights, whereas smaller $B$ stabilizes the solution at the cost of
additional shrinkage bias.

\paragraph{Selection-corrected regression.}
We then form the implied weights
\[
\widehat\omega_i=1+\widehat\zeta(X_i,Y_i),
\]
and estimate $\mu(t,x)=\E(Y\mid T=t,X=x)$ by a weighted nonparametric regression on complete cases.
The complexity of this final regression introduces an additional bias--variance trade-off: stronger
smoothing reduces variance but can attenuate heterogeneity in the estimated treatment effect, while
weaker smoothing can recover finer structure at the cost of increased variability when weights are
unstable. Finally,
\[
\widehat{\tau}_{t_1,t_0}(x)=\widehat\mu(t_1,x)-\widehat\mu(t_0,x).
\]

\paragraph{Tuning and inference.}
In practice, the first-stage smoothing levels, sieve dimensions, and regularization bound $B$ are
selected jointly to achieve a small integral equation residual and well-behaved implied weights,
avoiding solutions with many nonpositive or excessively large weights. Let $\pi_\textup{min}>0$ be a user-specified lower bound on the response probability.
In implementation, we enforce feasibility/bounded-positivity by truncating the implied weights to lie
in $[1,1/\pi_\textup{min}]$ (equivalently, truncating $\hat{\pi}$ to lie in $[\pi_\textup{min},1]$)
before the final weighted regression.
The complexity of the outcome
regression is selected separately. Once selected, all tuning parameters are held fixed across
bootstrap resamples for inference.

\section{Simulation details and additional results}\label{app:sim-more}

\subsection{Data generating process}
Binary variables are generated from Bernoulli distributions and continuous variables from Gaussian distributions. The covariate $X$ is generated marginally. The treatment $T$ is generated conditional on $X$ using either a logistic model (binary $T$) or a linear Gaussian model (continuous $T$). The outcome $Y$ is generated conditional on $(T,X)$ using either a logistic model (binary $Y$) or a linear Gaussian model (continuous $Y$), allowing for main effects of $T$ and $X$ and a $T\times X$ interaction.

Response indicators $(R^X,R^T,R^Y)$ follow logistic regression models with fixed slope coefficients. Intercepts are calibrated so that $\bP(R^\cdot=1)\approx0.8$. Outcome missingness is specified to align with Assumptions~\ref{ass1}--\ref{ass3} via
\[
\logit\,\bP(R^Y=1\mid\cdot)=\phi_0+0.4R^X+0.4R^T+u_a,\qquad a\in\{1,2,3\}.
\]

Table~\ref{tab:sim-params} lists all fixed slope coefficients and outcome model parameters. Intercepts $(\gamma_0,\eta_0,\phi_0)$ are calibrated to achieve the target missingness rates and are omitted.

\begin{table}[H]
\centering
\caption{Simulation parameter values.}
\label{tab:sim-params}
\setlength{\tabcolsep}{6pt}
\renewcommand{\arraystretch}{1.15}
\footnotesize
\begin{tabular}{p{0.18\linewidth} p{0.36\linewidth} p{0.42\linewidth}}
\toprule
Model & Binary case & Continuous case\\
\midrule
$X$ &
$p_X=0.5$ &
$(\mu_X,\sigma_X)=(0.2,1.0)$\\[3pt]

$T\mid X$ &
$(\alpha_0,\alpha_x)=(-0.3,0.9)$ &
$(\alpha_0,\alpha_x,\sigma_T)=(0.5,0.9,1.0)$\\[4pt]

$Y\mid(T,X)$ &
$(\beta_0,\beta_t,\beta_x,\beta_{tx})=(-0.4,1.1,0.9,0.5)$ &
$(\beta_0,\beta_t,\beta_x,\beta_{tx},\sigma_Y)=(-0.3,1.0,0.8,0.5,1.0)$\\[6pt]

$R^X\mid(X,T)$ &
$(\gamma_x,\gamma_t)=(0.6,-0.4)$ &
$(\gamma_x,\gamma_t)=(-1.0,0.6)$\\[3pt]

$R^T\mid(R^X,X,T)$ &
$(\eta_x,\eta_t,\eta_r)=(0.4,0.4,0.5)$ &
$(\eta_x,\eta_t,\eta_r)=(-0.6,-0.4,0.5)$\\[6pt]

$R^Y$ A1 ($u_1$) &
$u_1=-0.8X+0.9T$ &
$u_1=-0.4X-0.4T$\\
$R^Y$ A2 ($u_2$) &
$u_2=-0.8X+2.2Y$ &
$u_2=-0.4X-1.8Y$\\
$R^Y$ A3 ($u_3$) &
$u_3=0.9T+2.2Y$ &
$u_3=-0.4T-1.8Y$\\
\bottomrule
\end{tabular}
\end{table}

\subsection{Additional results under a null treatment effect.}

\begin{figure}[t]
    \centering
    \includegraphics[width=\linewidth]{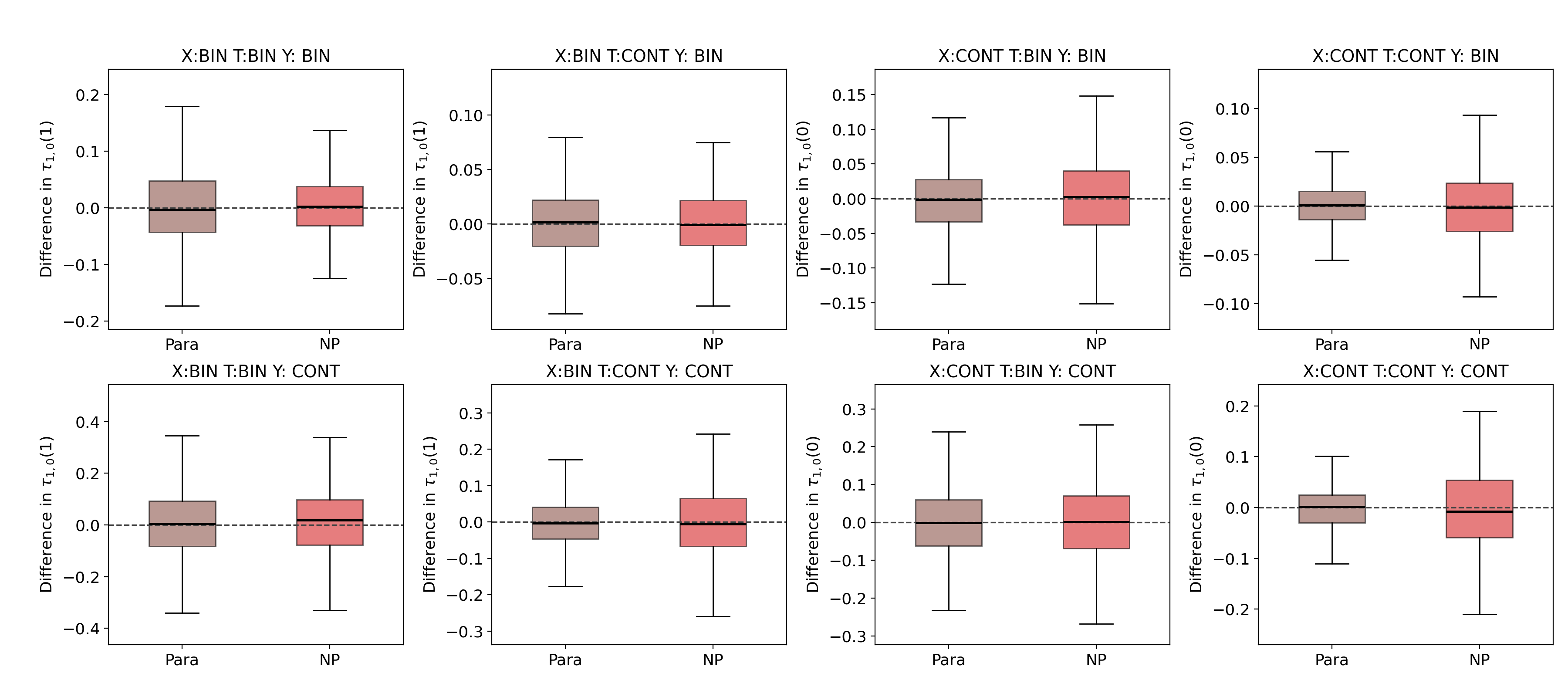}
\caption{Under a null effect ($\tau_{1,0}(x)=0$), both NP and Para remain centered near zero. The figure shows estimation error $\widehat{\tau}_{1,0}(x)-\tau_{1,0}(x)$ across eight data-type combinations under Assumption~\ref{ass2}; values closer to zero indicate better performance.}
    \label{fig:sim-null-bin}
\end{figure}

We consider a null effect variant of the simulation in which $\tau_{t_1,t_0}(x)=0$ for all $t_1,t_0,x$. This setting uses the same data generating process as the main simulation, except that the treatment effect parameters in the outcome model are set to zero, so that $Y \independent T \mid X=x$ for every $x$. As a result, the completeness condition in Theorem~\ref{the2} fails by construction. Because percent bias is not meaningful when $\tau_{t_1,t_0}(x)=0$, we assess performance using the estimation error $\widehat{\tau}_{t_1,t_0}(x)-\tau_{t_1,t_0}(x)$. In this null setting (Figure~\ref{fig:sim-null-bin}), both NP and Para remain centered near zero under Assumption~\ref{ass2}, providing empirical support for the identification result in part (ii) of Theorem~\ref{the2}.

\bibliographystyleSupp{apalike}
\bibliographySupp{ref}

\end{document}